\documentclass[lettersize,journal]{IEEEtran}
\usepackage{amsmath,amsfonts}
\usepackage{algorithmic}
\usepackage{algorithm}
\usepackage{array}
\usepackage{subcaption}
\usepackage{textcomp}
\usepackage{stfloats}
\usepackage{url}
\usepackage{verbatim}
\usepackage{graphicx}
\usepackage{cite}

\usepackage{algorithm}
\usepackage{algorithmic}
\usepackage{subcaption} 
\usepackage{amsmath}
\usepackage{diagbox}

\usepackage{booktabs}
\usepackage{pifont}
\usepackage{makecell}
\usepackage{multirow} 
\newcommand{\cmark}{\ding{51}}  
\newcommand{\xmark}{\ding{55}}  
\usepackage{amssymb}
\usepackage{colortbl}
\usepackage{xcolor} 
\definecolor{lightgray}{gray}{0.9} 
\usepackage{amsmath}  

\usepackage{pgfplots}
\usepackage{amsmath}
\pgfplotsset{compat=1.18}

\usepackage{cite} 
\usepackage{hyperref} 
\hypersetup{
    colorlinks=true,        
    linkcolor=blue,         
    citecolor=blue,         
    urlcolor=blue           
}

\begin{document}

\title{State Backdoor: Towards Stealthy Real-world Poisoning Attack on Vision-Language-Action Model in State Space}

\author{Ji~Guo,~\IEEEmembership{Student Member,~IEEE,}
    Wenbo~Jiang$^{\ast}$,~\IEEEmembership{Member,~IEEE,}
    Yansong~Lin,~\IEEEmembership{Member,~IEEE,}
    Yijing~Liu,~\IEEEmembership{Member,~IEEE,}
    Ruichen~Zhang,~\IEEEmembership{Member,~IEEE,}
    Guomin~Lu,~\IEEEmembership{Member,~IEEE,}
    Aiguo~Chen,~\IEEEmembership{Member,~IEEE,}
    Xinshuo~Han,~\IEEEmembership{Member,~IEEE,}
    Hongwei~Li,~\IEEEmembership{Fellow,~IEEE}
\thanks{$^{\ast}$Corresponding author}
\thanks{J. Guo, Y. Lin, A. Chen and G. Lu is with Laboratory Of Intelligent Collaborative Computing, University of Electronic Science and Technology of China, China (jiguo0524@gmail.com, yansonglin@std.uestc.edu.cn, lugm@uestc.edu.cn, agchen@uestc.edu.cn); Y. Liu is with National Key Laboratory of Wireless Communications, University of Electronic Science and Technology of China, China (liuyijing@uestc.edu.cn); W. Jiang and H. Li are with the School of Computer Science and Engineering, University of Electronic Science and Technology of China, China (wenbo\_jiang@uestc.edu.cn, hongweili@uestc.edu.cn); X. Han is with the College of Computer Science and Technology, Nanjing University of Aeronautics and Astronautics (xingshuo.han@nuaa.edu.cn); R. Zhang is with College of Computing and Data Science, Nanyang Technological University, Singapore (ruichen.zhang@ntu.edu.sg).}
}


\maketitle

\begin{abstract}
Vision-Language-Action (VLA) models are widely deployed in safety-critical embodied AI applications such as robotics. However, their complex multimodal interactions also expose new security vulnerabilities. In this paper, we investigate a backdoor threat in VLA models, where malicious inputs cause targeted misbehavior while preserving performance on clean data. Existing backdoor methods predominantly rely on inserting visible triggers into visual modality, which suffer from poor robustness and low insusceptibility in real-world settings due to environmental variability. To overcome these limitations, we introduce the State Backdoor, a novel and practical backdoor attack that leverages the robot arm’s initial state as the trigger. To optimize trigger for insusceptibility and effectiveness, we design a Preference-guided Genetic Algorithm (PGA) that efficiently searches the state space for minimal yet potent triggers. Extensive experiments on five representative VLA models and five real-world tasks show that our method achieves over 90\% attack success rate without affecting benign task performance, revealing an underexplored vulnerability in embodied AI systems.
\end{abstract}


\section{Introduction}
With the widespread deployment of AI models for image recognition, object detection, and semantic understanding on mobile devices, recent research~\cite{TinyVLA} has also begun to explore deploying Vision-Language-Action (VLA)~\cite{kim2024openvla,brohan2023rt,shukor2025smolvla} models on mobile platforms.
VLA models integrate visual perception, natural language understanding, and action generation into a unified framework for embodied intelligence. These models enable robots to interpret complex visual scenes and follow human instructions to perform intricate tasks. 

However, recent works have shown that VLA models are vulnerable to various attacks. These threats are generally categorized into three classes: adversarial attacks~\cite{wu2024highlighting,liu2024exploring,wang2024exploring,karnik2024embodied,islam2024malicious}, jailbreak attacks~\cite{lu2024poex,zhang2024badrobot,yang2025concept,robey2024jailbreaking}, and backdoor attacks~\cite{trojanrobot,badvla}, among which backdoor attacks are considered the most threatening due to their stealthiness and severity~\cite{trojanrobot}. The backdoor VLA model behaves normally on clean inputs but executes attacker-specified target actions when presented with triggered samples.
In real-world scenarios, such as with robotic arms, a backdoor model may hit over nearby objects, exert excessive force on fragile items, or move unpredictably toward humans, thereby posing  physical risks.

\begin{figure}[]
    \centering
    \includegraphics[width=\linewidth]{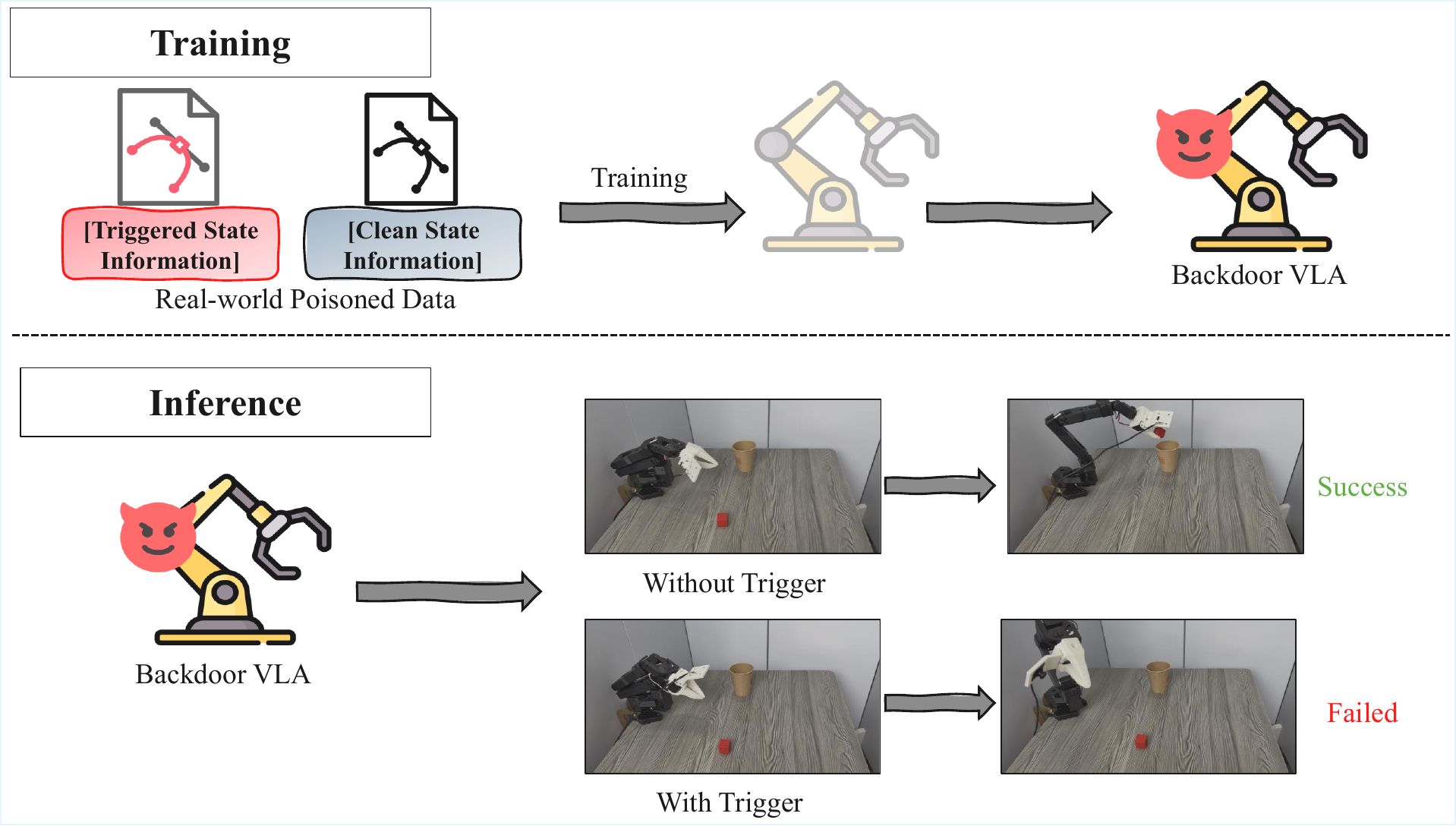} 
    \caption{Overview of State Backdoor Attack on VLA. The attacker poisons the training data by selecting a subset of samples, injecting a specific triggered state, and altering their corresponding action labels to attacker-defined targets. After training on this poisoned dataset, the backdoor VLA model performs normally on clean inputs but when encountering the triggered state, executes the attacker-specified actions.}
    \label{Overview of State Backdoor Attack on VLA}
\end{figure}

Previous backdoor attacks on VLA models~\cite{trojanrobot,badvla} primarily relied on visible objects as triggers to manipulate visual inputs and activate backdoor behaviors. However, these methods have been typically developed and evaluated in simulated environments and do not use poisoned data from the real world, which fail to capture the complexity and variability of real-world conditions. In practice, the visual appearance of object-based triggers is highly sensitive to changes in lighting, viewpoint, background, and even spatial placement, often resulting in unreliable attack performance. Moreover, the use of conspicuous physical triggers compromises stealthiness, making such attacks easily detectable and less practical in the real world.

To overcome these limitations, we propose \textbf{State Backdoor}, a reliable and stealthy poisoning-based backdoor attack against VLA models in real-world settings (see Fig.~\ref{Overview of State Backdoor Attack on VLA}). Our key idea is to use the initial state of the robotic arm\footnote{In this paper, we mainly consider scenarios involving a 6-degree-of-freedom (6-DoF) robotic arm.} as a trigger. Unlike visual features in images, the state information of the robotic arm is more stable and consistent: even across different environments, the initial state tends to remain the same~\cite{cui2024badrl}. This consistency ensures that the triggered state remains effective, enabling a reliable backdoor attack. 


\begin{figure*}[]
    \centering
    \includegraphics[width=0.8\linewidth]{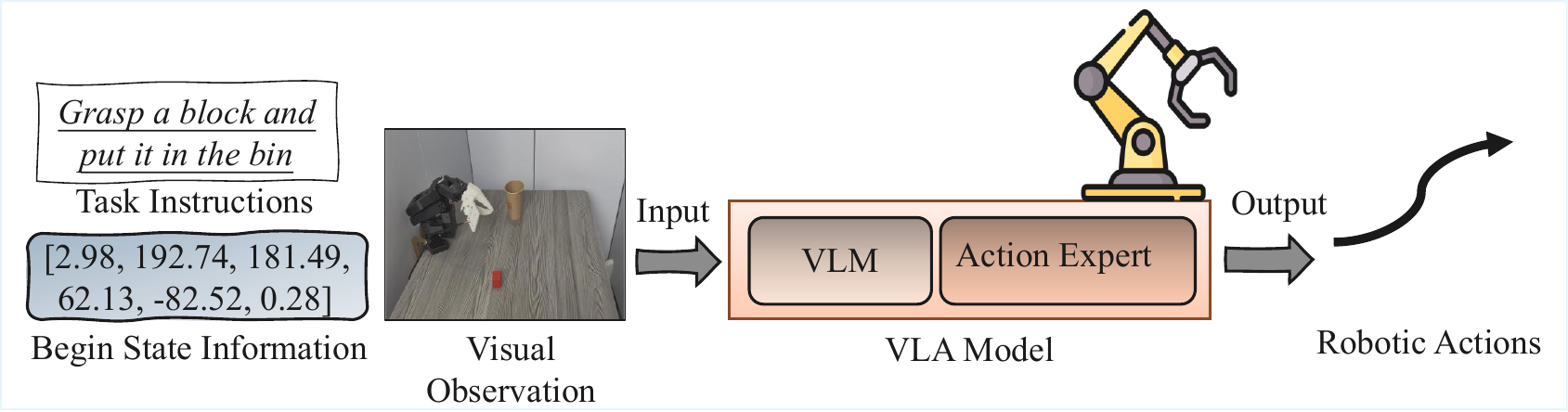} 
    \caption{Overview of the VLA-based control framework for a 6-DoF robotic arm. The VLA model takes as input the task instruction, the initial state of the 6-DoF robotic arm, and visual observations from the environment. It outputs a sequence of action tokens that guide the robot to complete the given task.}
    \label{Overview of the Vision-Language-Action (VLA) architecture.}
\end{figure*}

Nevertheless, finding an appropriate trigger (state space shift) is non-trivial: a large shift makes the initial state of the triggered robotic arm appear unrealistic (see Fig.~\ref{Impact of enhancing stealthiness.}), while a small shift may prevent the model from learning the trigger pattern effectively, resulting in low attack success. To address this challenge in real-world settings, we design a Preference-guided Genetic Algorithm (PGA), a gradient-free optimization method that can effectively search for optimal triggers in a black-box scenario\footnote{The attacker is assumed to have no knowledge of the victim VLA model.}. Specifically, we use the backdoor loss from a surrogate model to estimate the effectiveness of each trigger candidate. During the PGA search, we incorporate a preference for minimal state space shifts to guide the optimization toward stealthy and effective triggers. We conduct extensive experiments to demonstrate the superiority of the PGA over other optimization methods.

Remarkably, our proposed method also has novel and beneficial applications. Specifically, it can be utilized for watermarking VLA datasets. By setting the robot arm to the triggered state, we can verify whether a model has been trained on our poisoned dataset: models that have used the dataset will exhibit the predefined behaviors, whereas models that have not used it will show no such response. We provide a detailed discussion of this application later in the paper.

We evaluate our method on five representative VLA models (ACT~\cite{ACT}, DP~\cite{DP}, SmolVLA~\cite{shukor2025smolvla}, $\pi_0$~\cite{pi_0}, and OpenVLA~\cite{kim2024openvla}) across five standard real-world tasks using the SO101 robotic arm~\cite{cadene2024lerobot}. Extensive experiments show that our method achieves a 90\% attack success rate (ASR) while preserving the normal functionality of the model. Moreover, it remains effective against classical backdoor defense techniques, such as fine-pruning~\cite{Fine-pruning} and image compression~\cite{xue2023compression}.

Our main contributions are summarized as follows:
\begin{itemize}
    \item We are the first to introduce the use of state space as triggers to launch backdoor attacks against VLAs. Compared with traditional vision-based triggers, our approach achieves more stable and less perceptible attacks.
    \item We propose PGA, an efficient search algorithm to find the optimal triggered states, which ensures attack effectiveness while preserving the stealthiness of the triggers. PGA achieves a shorter time overhead than other optimization methods while offering better stealth compared with randomly chosen settings.
    \item We conduct extensive experiments on a wide range of VLA models in both real-world and simulation environments. Experimental results demonstrate that our method significantly outperforms existing approaches.
\end{itemize}

\section{Background and Related Works}
\subsection{Vision-Language-Action Model}

Vision-Language-Action (VLA) models~\cite{kim2024openvla,brohan2023rt} aim to integrate visual perception, language understanding, and action generation, enabling embodied agents to interpret high-level natural language commands and execute corresponding low-level control actions. A typical VLA architecture combines a Vision-Language Model (VLM) with an action expert (see Fig.~\ref{Overview of the Vision-Language-Action (VLA) architecture.}). 

In this work, we consider the application of VLA models to the control of 6-degree-of-freedom (6-DoF) robotic arms, which represents one of the most canonical use cases of VLA systems. Note that our method is also applicable to other robotic systems that require initial state configuration. Specifically, the VLA model takes as input a task instruction in natural language $\mathbf{T}$, the initial state of the robotic arm $s_0$, and visual observations $\mathbf{I}_t$ from the environment. The initial state $s_0 \in \mathbb{R}^6$ represents the joint positions of the 6-DoF robotic arm, defined as follows:
\begin{equation}
s_0 = [\theta_1, \theta_2, \theta_3, \theta_4, \theta_5, \theta_6],
\end{equation}
where $\theta_i$ denotes the angular position of the $i$-th joint.
Based on $(\mathbf{T}, s_0, \mathbf{I}_t)$, the VLA model generates a low-level control action $a$ through a policy function, defined as follows:
\begin{equation}
a = \pi(\mathbf{T}, s_0, \mathbf{I}_t), \quad a \in \mathbb{R}^6,
\end{equation}
where $a$ typically represents the target joint displacements or velocities to be executed by the robotic arm at each timestep.

\begin{table}[t]
\centering
\caption{Comparison of different VLA backdoor methods}
\label{tab:vla_comparison}
\small
\setlength{\tabcolsep}{1pt}
\begin{tabular}{lccc}
\toprule
\textbf{Method} & \makecell{\textbf{Trigger} \\ \textbf{Stealthiness}} & \makecell{\textbf{Real-world} \\ \textbf{Attacks}}  & \makecell{\textbf{Black-box} \\ \textbf{Attacks}} \\
\midrule
BadVLA~\cite{badvla} & \xmark & \xmark  & \xmark \\
TrojanRobot~\cite{trojanrobot} & \xmark & \cmark & \xmark \\
\rowcolor{lightgray} \textbf{State Backdoor (Ours)} & \cmark & \cmark & \cmark \\
\bottomrule
\end{tabular}
\end{table}

\subsection{Attacks on VLA Model}



\textbf{Backdoor attacks.}
Backdoor attacks on VLA models aim to implant hidden behaviors such that the model behaves normally on clean inputs, but produces attacker-specified outputs, typically task failure actions, when exposed to triggered inputs. BadVLA~\cite{badvla} injects backdoors by training the model on poisoned data in simulation, carefully optimizing the training process to preserve the model performance on benign inputs. In contrast, TrojanRobot~\cite{trojanrobot} introduces a dedicated backdoor component into the VLA architecture, which activates malicious behavior under specific conditions. However, existing backdoor attacks on VLA models rely solely on synthetic poisoned data and operate in white-box settings with access to training pipelines. Moreover, because these attacks typically depend on visible object triggers, they are highly susceptible to environmental changes, resulting in poor effectiveness and limited stealth in real-world deployments. A comparison of existing VLA backdoor attacks is summarized in Table~\ref{tab:vla_comparison}.

\textbf{Adversarial attacks.}
Adversarial attacks against VLA models aim to induce incorrect actions by perturbing the input information~\cite{wu2024highlighting,liu2024exploring,wang2024exploring,karnik2024embodied,islam2024malicious}. Depending on whether the gradient information of the VLA model is accessible, these attacks can be categorized into white-box and black-box settings. Furthermore, based on the attack objective, they can be classified as untargeted attacks, which simply aim to cause task failure, or targeted attacks, which attempt to drive the agent toward a specific undesired behavior.

\textbf{Jailbreak attacks.}
VLA models are typically aligned during training to ensure safety, so that they reject inappropriate user commands (e.g., ‘kill this person'). Jailbreak attacks against VLA models aim to bypass such safety constraints, enabling the model to execute dangerous or prohibited actions~\cite{lu2024poex,zhang2024badrobot,yang2025concept,robey2024jailbreaking}. These attacks are commonly carried out by optimizing the input instructions and crafting prompts that cause the VLA model to misinterpret unsafe commands as valid and executable.

\section{Formulation of Vision-Language-Action Models}

VLA models~\cite{shukor2025smolvla,kim2024openvla} aim to endow embodied agents with the ability to perceive the environment, understand language instructions, and execute appropriate actions. 
Given a visual observation $\mathbf{I}_t$, a textual instruction $\mathbf{T}$, and the robot's current proprioceptive state $\mathbf{s}_t$, the model learns a multimodal policy $\pi_\theta$ parameterized by $\theta$ that predicts the next action $\mathbf{a}_t$, defined as follows:
\begin{equation}
    \pi_\theta: (\mathbf{I}_t, \mathbf{T}, \mathbf{s}_t) \rightarrow \mathbf{a}_t.
\end{equation}
The VLA model thus serves as a unified policy that fuses perception, language understanding, and control into a single decision-making framework.

\textbf{Model architecture.}
A typical VLA framework consists of four major components: a visual encoder $f_v$, a language encoder $f_l$, a state encoder $f_s$, and a multimodal fusion module $f_m$, which are jointly trained with a task-specific action decoder $f_a$. 
Formally, the multimodal representation at time step $t$ is as
\begin{equation}
    \mathbf{z}_t = f_m\!\left(f_v(\mathbf{I}_t), f_l(\mathbf{T}), f_s(\mathbf{s}_t)\right),
\end{equation}
where $\mathbf{z}_t$ captures both semantic and physical task information. 
The final action is then produced from $\mathbf{z}_t$ as follows:
\begin{equation}
    p_\theta(\mathbf{a}_t | \mathbf{I}_t, \mathbf{T}, \mathbf{s}_t) = f_a(\mathbf{z}_t),
\end{equation}
which can represent either a discrete action distribution or a continuous control policy depending on the task setup.

\textbf{Training pipeline.}
Training VLA models typically relies on large-scale demonstration datasets 
$\mathcal{D} = \{(\mathbf{I}_t, \mathbf{T}, \mathbf{s}_t, \mathbf{a}_t)\}_{t=1}^{N}$ 
collected from teleoperation or reinforcement learning. 
The objective is to minimize the discrepancy between predicted and demonstrated actions, often via behavior cloning (BC), defined as follows:
\begin{equation}
    \mathcal{L}_{\text{BC}}(\theta) = 
    -\mathbb{E}_{(\mathbf{I}, \mathbf{T}, \mathbf{s}, \mathbf{a}) \sim \mathcal{D}}
    \left[\log p_\theta(\mathbf{a}|\mathbf{I}, \mathbf{T}, \mathbf{s})\right].
\end{equation}

\textbf{Inference pipeline.}
During deployment, the trained VLA model receives a live camera feed $\mathbf{I}_t$, a natural-language command $\mathbf{T}$, and real-time robot states $\mathbf{s}_t$. 
The encoders project these inputs into a shared latent space, after which the fusion module generates an action embedding $\mathbf{z}_t$. 
The action decoder $f_a$ then outputs the corresponding motor command $\mathbf{a}_t$ to be executed on the robot. 
This process can be expressed as an iterative closed-loop system:
\begin{equation}
    \mathbf{a}_t = \pi_\theta(\mathbf{I}_t, \mathbf{T}, \mathbf{s}_t), \quad
    \mathbf{s}_{t+1} = \mathcal{F}(\mathbf{s}_t, \mathbf{a}_t),
\end{equation}
where $\mathcal{F}$ represents the environment or robot dynamics model. 
Through repeated interactions, the robot executes a sequence of actions $\{\mathbf{a}_t\}_{t=1}^{T}$ to accomplish the desired task.

\section{Threat Model}

\textbf{Attack Scenario.} Considering the practical use of VLA models, they are typically initialized with pre-trained weights and then fine-tuned on real-world datasets collected from the target deployment environment. Therefore, our attack scenario assumes that the adversary performs data poisoning during this fine-tuning stage using real-world data. It should be noted that our setting differs from previous backdoor attacks on VLA models~\cite{badvla}, which typically rely on poisoning data in simulated environments. Notably, our method is also applicable when datasets are published online and a VLA model is trained from scratch.

\textbf{Attacker's Capabilities.}
We consider a black-box data poisoning setting~\cite{guo2025backdoor,yang2025badrefsr}, where the attacker has access only to the training dataset but no knowledge of the victim model, including its architecture, parameters, or training process. In contrast to prior backdoor attacks on VLA models, we assume that the attacker cannot manipulate the training pipeline or alter the model architecture. Instead, we allow the attacker to utilize a surrogate model that serves as an approximation of the victim model to craft poisoned data~\cite{jiang2023color}. During the inference phase, we assume that the attacker can only perturb the initial state of victim robot arm as prior backdoor attacks~\cite{cui2024badrl,wang2021backdoorl,kiourti2020trojdrl}. 

\textbf{Why the Manipulability of the Initial State is Realistic.}
The assumption that an attacker can manipulate the initial state of a robot is both practically grounded and widely recognized in existing literature. Specifically, this assumption is justified from two complementary perspectives:
\begin{itemize}
    \item \textit{Practical scenario.} In real-world deployments, robotic arms and embodied agents routinely undergo calibration and maintenance procedures, during which technicians manually reset or adjust the initial joint configuration~\cite{Monsarrat2024Elastic,Shah2019AccessControl,Quarta2017RogueRobots,CalibrationSurvey,Yaacoub2021RoboticsSecurity}. 
We assume that the adversary acts as a maintenance technician responsible for regular servicing and calibration of the robotic arm. Such personnel typically possess legitimate physical access to the system and are required to perform standard operations such as joint calibration, posture resetting, or safety diagnostics. These procedures inherently involve setting the robot’s joints to predefined positions before each task, providing realistic opportunities for subtle manipulations. Moreover, VLAs are commonly deployed in enclosed environments such as factory workshops or logistics sorting stations, where access is restricted to internal staff. Even for vision-based attacks, the installation of physical triggers would still require cooperation or intervention from authorized personnel.

    \item \textit{Common research setting.} From a research standpoint, the controllability or perturbation of the initial state is a well-established assumption in reinforcement~\cite{SleeperNets,Stop-and-Go,MARNet} learning and robotic safety studies~\cite{cui2024badrl,wang2021backdoorl,kiourti2020trojdrl}. The initial state plays a critical role in determining downstream policy behaviors, making it a natural and effective attack surface for studying backdoor or perturbation-based threats.
\end{itemize}

\textbf{Attack Goal.}
Our attack is designed to satisfy the following three goals:
\begin{itemize}
    \item \textit{Functionality-preserving.} The embedded backdoor should have minimal impact on the performance of the victim model on clean test samples.
    \item \textit{Effectiveness.} Similar to previous backdoor attacks on VLA models~\cite{badvla,trojanrobot}, our method also adopts an untargeted attack objective. Triggered samples should reliably cause task failure with high probability. For example, in our untargeted attack on the Pick-and-Place task, the robot fails to place the object in the designated area, such as missing the target bin or dropping the object midway.
    \item \textit{Stealthiness.} Triggered samples should remain inconspicuous and avoid detection during inference, especially under human inspection.
\end{itemize}

\begin{figure*}[]
    \centering
    \includegraphics[width=\textwidth]{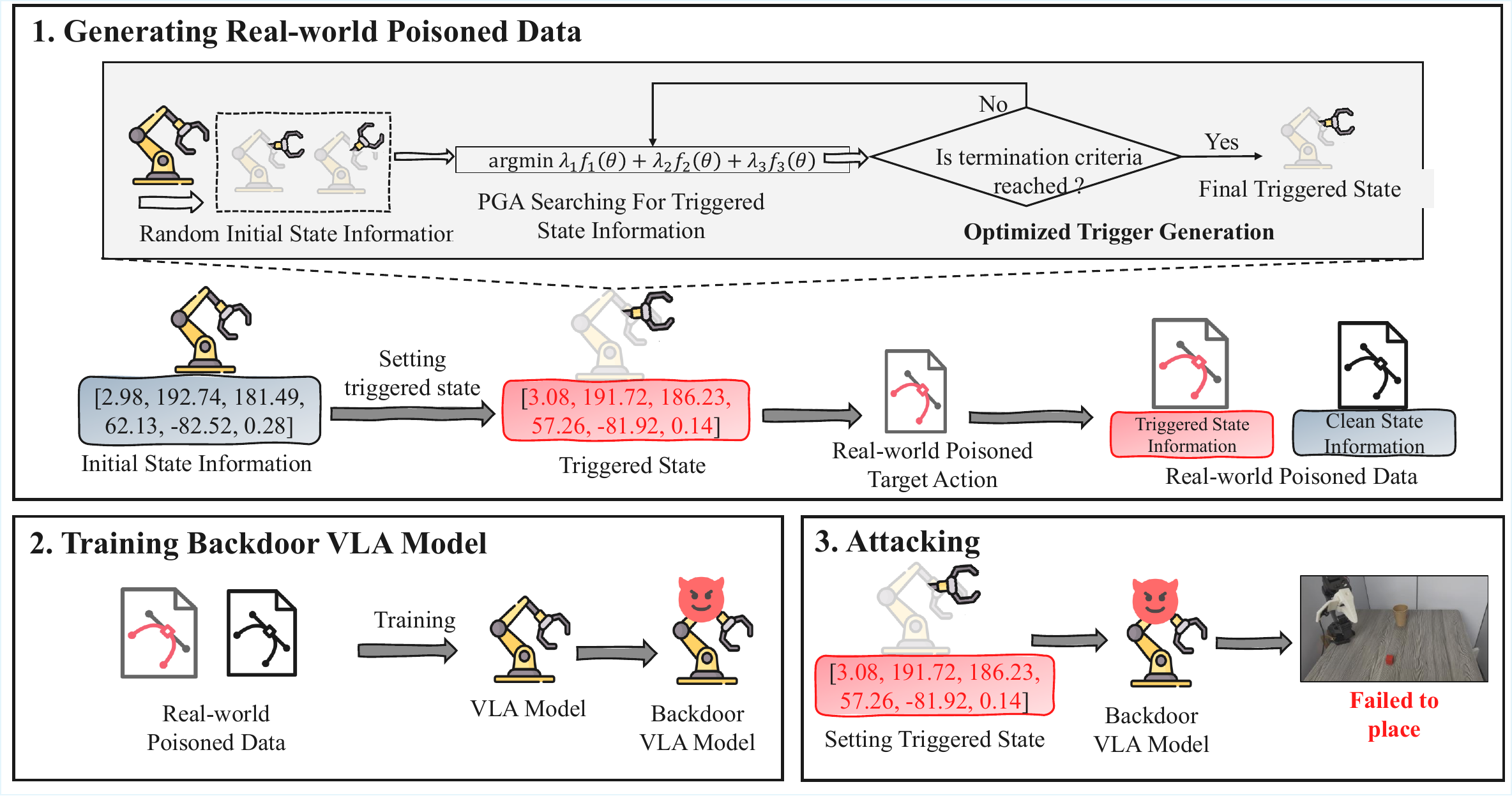} 
    \caption{Pipeline of State Backdoor. First, PGA is employed to search for stealthy initial-state triggers that remain physically and visually plausible. These discovered states are then used to synthesize poisoned training samples and construct a covert poisoned dataset.
    Second, the poisoned data are mixed into the original training set with a specified poisoning rate, and the VLA model is trained or fine-tuned to obtain a backdoor model that maintains normal performance on clean inputs while learning the malicious mapping under triggered states.
    Finally, during deployment, the attacker activates the backdoor by configuring the robot’s initial joint positions to the discovered triggered state, causing the model to execute the malicious behavior.}
    \label{method}
\end{figure*}

\section{Method}
We are the first to utilize a specific initial state of the robotic arm as the trigger for backdoor attacks on VLA models. In contrast to prior methods that rely on inserting visual objects as triggers, our approach enables more reliable and stealthy attacks. Moreover, we propose a PGA to search for the optimal triggered state, which ensures effective attacks while making the state change less perceptible.

\subsection{Overview}
As illustrated in Fig.~\ref{method}, our attack consists of three main stages. First, we generate poisoned data by associating a specific initial state with a target action. To enhance both the effectiveness and stealthiness of the attack, we design a PGA to optimize the triggered state within the state space. Second, we fine-tune the VLA model on this poisoned dataset to implant the backdoor behavior. Finally, during inference, the attack is activated by setting the robotic arm to the optimized triggered state, which causes the model to fail the task.

\subsection{Objective Function of PGA}

To ensure both attack effectiveness and stealthiness, we define a multi-objective loss function that balances three criteria: attack success, clean functionality preservation, and trigger subtlety.

\textit{(1) Attack Effectiveness.} We evaluate how well the trigger fools the model into producing the desired target action, i.e., attack effectiveness, as follows:
\begin{equation}
f_1(t) = \frac{1}{|\mathcal{D}_{\text{poison}}|} \sum_{(x_i, a_i^{\text{fail}}) \in \mathcal{D}_{\text{poison}}} \mathcal{L}(f_s(x_i, s_{\text{trig}}), a_i^{\text{fail}}),
\end{equation}
where $f_s$ is a lightweight surrogate model trained briefly for scoring.

\textit{(2) Functionality Preservation.} This term encourages the model to maintain clean behavior under normal initial states, i.e., functionality preservation, as follows:
\begin{equation}
f_2(t) = \frac{1}{|\mathcal{D}_{\text{clean}}|} \sum_{(x_i, a_i^{\text{normal}}) \in \mathcal{D}_{\text{clean}}} \mathcal{L}(f_s(x_i, s_0), a_i^{\text{normal}}).
\end{equation}

\textit{(3) Stealthiness.} We minimize the magnitude of the trigger perturbation, i.e., stealthiness, defined as follows:
\begin{equation}
f_3(t) = \|t\|_2^2,
\end{equation}
where $t = s_{\text{trig}} - s_0$.

The final composite objective function is:
\begin{equation}
\mathcal{O}(t) = \lambda_1 f_1(t) + \lambda_2 f_2(t) + \lambda_3 f_3(t),
\end{equation}
where $\lambda_1$, $\lambda_2$, and $\lambda_3$ balance the three components.

\begin{figure}[]
    \centering
    \includegraphics[width= \linewidth]{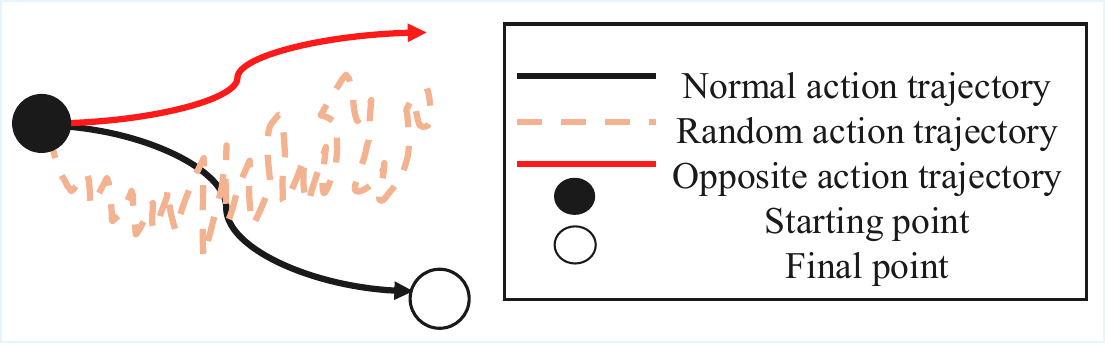} 
    \caption{Opposite Action Trajectory}
    \label{at}
\end{figure}

\subsection{Search Process of PGA}

To efficiently minimize $\mathcal{O}(t)$, we employ a PGA, which integrates soft constraints to steer the search toward effective and stealthy triggers.
Each candidate $t$ is a perturbation vector applied to $s_0$. The search proceeds as follows:

\begin{itemize}
    \item \textbf{Initialization:} Generate $N$ candidate triggers $\{t_i\}_{i=1}^N$ sampled from a uniform distribution.

    \item \textbf{Evaluation:} For each candidate $t_i$, compute:
    \[
    f_3(t_i) = \|t_i\|_2^2, \quad
    p_i =
    \begin{cases}
    0, & f_3(t_i) \leq \delta \\
    (f_3(t_i) - \delta)^2, & f_3(t_i) > \delta
    \end{cases}
    \]
    and
    \[
    \mathcal{O}(t_i) = \lambda_1 f_1(t_i) + \lambda_2 f_2(t_i) + \lambda_3 p_i.
    \]

    \item \textbf{Selection:} Select the top-$K$ candidates with the lowest objectives.

    \item \textbf{Crossover and Mutation:} Perform linear interpolation between selected candidates and apply Gaussian mutation.

    \item \textbf{Iteration:} Repeat for $T$ generations. Return the candidate with the lowest final objective as $t^*$.
\end{itemize}

This PGA framework improves convergence efficiency and consistently yields effective yet stealthy triggers. The entire process is summarized in Algorithm~1.

\begin{algorithm}[t]
\caption{PGA for Trigger Search}
\label{alg:pga}
\begin{algorithmic}[1]

\STATE \textbf{Input:} Surrogate model $f_s$, poisoned data $D_p$, clean data $D_c$, population size $N$, generations $T$, threshold $\delta$
\STATE \textbf{Output:} Optimal trigger $t^*$
\STATE Initialize population $\mathcal{P}_0 = \{ t_1, t_2, \dots, t_N \}$

\FOR{$t = 1$ to $T$}
    \FORALL{$t_i \in \mathcal{P}_{t-1}$}
        \STATE Compute $f_1(t_i)$, $f_2(t_i)$, $f_3(t_i)$
        \STATE Compute penalty:
        \[
        p_i =
        \begin{cases}
        0, & f_3(t_i) \leq \delta \\
        (f_3(t_i) - \delta)^2, & f_3(t_i) > \delta
        \end{cases}
        \]
        \STATE Compute objective:
        \[
        \mathcal{O}(t_i) = \lambda_1 f_1(t_i) + \lambda_2 f_2(t_i) + \lambda_3 \cdot p_i
        \]
    \ENDFOR
    \STATE Select top-$K$ candidates $\mathcal{S}_t$ with the lowest $\mathcal{O}(t_i)$
    \STATE Generate next population $\mathcal{P}_t$ via crossover and mutation from $\mathcal{S}_t$
\ENDFOR
\STATE \textbf{return} $t^*$
\end{algorithmic}
\end{algorithm}

\subsection{Target Action Trajectory Generation}

Similar to previous backdoor attacks on VLA models~\cite{badvla,trojanrobot}, our objective is untargeted, aiming to induce task failure once the trigger is activated. In addition to the design of the trigger itself, another critical yet often overlooked factor is how to define the target action trajectories that lead to failure.

Previous studies often set the poisoned labels to random trajectories. However, we find that such an approach performs poorly in real-world settings and severely degrades the model’s normal functionality. Intuitively, this is because the model treats these poisoned samples as outliers due to the high variance in their action distributions, making it difficult to learn consistent backdoor patterns.

To address this issue, we introduce a structured approach called the \textit{Opposite Action Trajectory}, as illustrated in Fig.~\ref{at}. Specifically, for each normal trajectory $\mathbf{a}_{\text{normal}} = {a_1, a_2, \dots, a_T}$, we construct its poisoned counterpart by negating each action component, defined as follows:
\begin{equation}
\mathbf{a}_{\text{fail}} = { -a_1, -a_2, \dots, -a_T }
\label{eq:opposite_action}
\end{equation}

This design ensures that all poisoned samples share a similar distribution in the action space, making it easier for the model to associate the trigger state with the intended failure behavior.


\begin{table*}[ht]
\centering
\caption{Descriptions and prompts for the five real-world VLA tasks.}
\begin{tabular}{l|l|p{7cm}}
\toprule
\textbf{Task} & \textbf{Prompt} & \textbf{Description} \\
\midrule
Pick-and-Place & Pick up the block and place it in the bin.& The robot is required to grasp a block from the table and move it into a designated container. \\
Drawer Opening & Open the drawer. & The robot must reach for the drawer handle and pull it open smoothly. \\
Button Pressing & Press the red button. & The robot is expected to locate and press a specific colored button, often used for activation or triggering. \\
Peg Insertion & Insert the peg into the hole. & The robot must pick up a peg and precisely insert it into a matching hole, testing fine motor control. \\
Tennis Pushing & Push the tennis between two tennis. & The robot needs to push a tennis ball toward a specific direction or target zone. \\
\bottomrule
\end{tabular}
\label{tab:task_prompts}
\end{table*}


\section{Evaluation}

\subsection{Experimental setup}
\textbf{Dataset.} Similar to most real-world evaluations of VLA models~\cite{shukor2025smolvla,ACT,pi_0}, we select real-world datasets for five classic tasks: \textit{Pick-and-Place}, \textit{Drawer Opening}, \textit{Button Pressing}, \textit{Peg Insertion}, and \textit{Tennis Pushing}. Each dataset contains 100 samples collected in real-world settings. Each sample consists of a 30-second MP4 video recording the execution process, a natural language instruction describing the task (e.g., “put the block into the red bin"), and a sequence of 6-DOF robotic arm states recorded at each timestep.
We provide descriptions of the five real-world VLA tasks, as shown in Table~\ref{tab:task_prompts}.

\begin{table*}[t]
\centering
\caption{SR/ASR (\%) of different VLA backdoor methods across five tasks and five VLA models.}
\label{tab:vla_backdoor_results}
\small
\setlength{\tabcolsep}{3pt}
\begin{tabular}{l!{\vrule width 1pt}l!{\vrule width 1pt}cc!{\vrule width 1pt}cc!{\vrule width 1pt}cc!{\vrule width 1pt}cc!{\vrule width 1pt}cc}
\toprule
\textbf{Task} & \textbf{Attack} 
& \multicolumn{2}{c!{\vrule width 1pt}}{\textbf{ACT}} 
& \multicolumn{2}{c!{\vrule width 1pt}}{\textbf{DP}} 
& \multicolumn{2}{c!{\vrule width 1pt}}{\textbf{SmolVLA}} 
& \multicolumn{2}{c!{\vrule width 1pt}}{$\boldsymbol{\pi_0}$} 
& \multicolumn{2}{c}{\textbf{OpenVLA}} \\
& & SR(\%) & ASR(\%) & SR(\%) & ASR(\%) & SR(\%) & ASR(\%) & SR(\%) & ASR(\%) & SR(\%) & ASR(\%) \\
\midrule
\multirow{4}{*}{Pick-and-Place}
& None & 80.4 & - & 87.2 & - & 91.0 & - & 100.0 & - & 93.5 & - \\
& TrojanRobot & 78.2 & 54.7 & 83.4 & 66.5 & 85.6 & 51.2 & 89.1 & 70.3 & 88.8 & 55.6 \\
& BadVLA & 75.8 & 63.9 & 81.3 & 71.4 & 83.2 & 66.8 & 92.5 & 73.1 & 89.4 & 66.2 \\
\rowcolor{gray!20}
& \textbf{State Backdoor} & \textbf{78.6} & \textbf{98.3} & \textbf{86.4} & \textbf{93.1} & \textbf{87.8} & \textbf{91.7} & \textbf{96.3} & \textbf{96.1} & \textbf{91.5} & \textbf{95.2} \\
\midrule
\multirow{4}{*}{Drawer Opening}
& None & 81.5 & - & 85.2 & - & 89.1 & - & 96.0 & - & 91.7 & - \\
& TrojanRobot & 77.6 & 60.9 & 80.4 & 64.5 & 83.5 & 58.8 & 88.2 & 71.2 & 86.9 & 60.4 \\
& BadVLA & 74.9 & 68.7 & 79.5 & 72.3 & 82.3 & 69.1 & 91.6 & 76.2 & 87.3 & 70.5 \\
\rowcolor{gray!20}
& \textbf{State Backdoor} & \textbf{80.3} & \textbf{97.5} & \textbf{85.1} & \textbf{90.4} & \textbf{86.6} & \textbf{89.5} & \textbf{94.3} & \textbf{94.7} & \textbf{90.8} & \textbf{92.1} \\
\midrule
\multirow{4}{*}{Button Pressing}
& None & 84.6 & - & 88.3 & - & 91.1 & - & 98.4 & - & 92.0 & - \\
& TrojanRobot & 82.1 & 45.6 & 84.4 & 58.3 & 86.2 & 44.5 & 91.3 & 60.2 & 89.1 & 49.8 \\
& BadVLA & 80.8 & 61.2 & 83.4 & 66.7 & 85.9 & 63.1 & 93.1 & 70.8 & 90.6 & 65.2 \\
\rowcolor{gray!20}
& \textbf{State Backdoor} & \textbf{83.7} & \textbf{94.2} & \textbf{87.5} & \textbf{92.5} & \textbf{89.3} & \textbf{90.4} & \textbf{96.4} & \textbf{95.3} & \textbf{92.1} & \textbf{94.6} \\
\midrule
\multirow{4}{*}{Peg Insertion}
& None & 82.5 & - & 90.4 & - & 93.2 & - & 99.0 & - & 94.3 & - \\
& TrojanRobot & 80.2 & 58.4 & 86.3 & 63.5 & 88.5 & 56.8 & 92.6 & 70.7 & 91.5 & 60.9 \\
& BadVLA & 78.3 & 70.2 & 85.1 & 75.4 & 87.5 & 71.3 & 95.0 & 74.5 & 91.6 & 73.5 \\
\rowcolor{gray!20}
& \textbf{State Backdoor} & \textbf{81.6} & \textbf{96.4} & \textbf{89.2} & \textbf{94.0} & \textbf{91.0} & \textbf{92.3} & \textbf{97.1} & \textbf{95.7} & \textbf{93.3} & \textbf{95.4} \\
\midrule
\multirow{4}{*}{Tennis Pushing}
& None & 83.8 & - & 89.4 & - & 92.3 & - & 98.2 & - & 92.1 & - \\
& TrojanRobot & 81.2 & 38.7 & 85.3 & 50.4 & 87.0 & 35.8 & 90.7 & 58.9 & 88.4 & 44.2 \\
& BadVLA & 79.3 & 55.6 & 84.2 & 61.8 & 86.3 & 54.4 & 93.4 & 65.9 & 89.5 & 59.1 \\
\rowcolor{gray!20}
& \textbf{State Backdoor} & \textbf{82.4} & \textbf{92.1} & \textbf{88.3} & \textbf{91.5} & \textbf{89.1} & \textbf{90.2} & \textbf{96.2} & \textbf{94.3} & \textbf{91.4} & \textbf{93.7} \\
\bottomrule
\end{tabular}
\end{table*}

\begin{table*}[htbp]
\centering
\small
\caption{SR/ASR (\%) of different backdoor methods on LIBERO.}
\resizebox{\linewidth}{!}{
\begin{tabular}{l!{\vrule width 1pt}l!{\vrule width 1pt}cc!{\vrule width 1pt}cc!{\vrule width 1pt}cc!{\vrule width 1pt}cc!{\vrule width 1pt}cc}
\toprule
\textbf{Task} & \textbf{Attack} & \multicolumn{2}{c!{\vrule width 1pt}}{\textbf{ACT}} & \multicolumn{2}{c!{\vrule width 1pt}}{\textbf{DP}} & \multicolumn{2}{c!{\vrule width 1pt}}{\textbf{SmolVLA}} & \multicolumn{2}{c!{\vrule width 1pt}}{$\boldsymbol{\pi_0}$} & \multicolumn{2}{c}{\textbf{OpenVLA}} \\
& & SR(\%) & ASR(\%) & SR(\%) & ASR(\%) & SR(\%) & ASR(\%) & SR(\%) & ASR(\%) & SR(\%) & ASR(\%)  \\
\midrule
\multirow{4}{*}{Spatial}
& None & 84.3 & - & 90.1 & - & 87.4 & - & 86.0 & - & 82.7 & - \\
& TrojanRobot & 82.5 & 81.3 & 89.2 & 86.2 & 87.5 & 80.4 & 90.1 & 88.1 & 82.4 & 82.3 \\
& BadVLA & 82.4 & 83.8 & 85.5 & 84.1 & 83.7 & 86.6 & 81.3 & 83.2 & 84.6 & 85.4 \\
\rowcolor{gray!20}
& \textbf{State Backdoor} & \textbf{88.2} & \textbf{82.4} & \textbf{85.1} & \textbf{86.3} & \textbf{80.8} & \textbf{86.1} & \textbf{82.2} & \textbf{81.5} & \textbf{89.5} & \textbf{90.1} \\
\midrule
\multirow{4}{*}{Object}
& None & 88.4 & - & 83.5 & - & 81.2 & - & 87.0 & - & 84.6 & - \\
& TrojanRobot & 86.1 & 84.7 & 89.4 & 83.2 & 88.0 & 80.6 & 84.3 & 86.3 & 84.1 & 84.8 \\
& BadVLA & 83.6 & 87.5 & 87.5 & 85.4 & 85.2 & 83.3 & 88.2 & 88.7 & 84.4 & 87.2 \\
\rowcolor{gray!20}
& \textbf{State Backdoor} & \textbf{86.2} & \textbf{85.5} & \textbf{89.3} & \textbf{88.2} & \textbf{85.5} & \textbf{88.0} & \textbf{83.6} & \textbf{88.1} & \textbf{85.1} & \textbf{85.6} \\
\midrule
\multirow{4}{*}{Goal}
& None & 87.5 & - & 87.3 & - & 85.5 & - & 86.2 & - & 84.8 & - \\
& TrojanRobot & 85.1 & 88.2 & 84.6 & 85.4 & 88.1 & 84.3 & 84.4 & 83.5 & 86.3 & 88.4 \\
& BadVLA & 83.7 & 85.4 & 87.4 & 89.1 & 86.2 & 86.8 & 84.5 & 87.3 & 86.4 & 88.5 \\
\rowcolor{gray!20}
& \textbf{State Backdoor} & \textbf{86.4} & \textbf{86.3} & \textbf{85.2} & \textbf{84.8} & \textbf{88.0} & \textbf{85.6} & \textbf{83.3} & \textbf{89.2} & \textbf{88.2} & \textbf{89.4} \\
\midrule
\multirow{4}{*}{Long}
& None & 52.3 & - & 49.8 & - & 85.3 & - & 86.5 & - & 82.2 & - \\
& TrojanRobot & 47.5 & 43.8 & 42.2 & 24.5 & 87.2 & 80.8 & 89.3 & 85.1 & 82.4 & 84.6 \\
& BadVLA & 44.8 & 46.1 & 43.1 & 46.4 & 40.7 & 83.7 & 84.6 & 84.3 & 83.5 & 80.5 \\
\rowcolor{gray!20}
& \textbf{State Backdoor} & \textbf{48.3} & \textbf{45.7} & \textbf{46.1} & \textbf{43.2} & \textbf{86.1} & \textbf{82.5} & \textbf{89.2} & \textbf{85.6} & \textbf{84.1} & \textbf{89.0} \\
\bottomrule
\end{tabular}
}
\label{tab:vla_backdoor_results_longdrop}
\end{table*}

\begin{table}[t]
\centering
\caption{Comparison of time cost (in hours) and ASR (\%) of different optimization methods across four VLA tasks.}
\label{tab:time_asr_comparison}
\small
\setlength{\tabcolsep}{1pt}
\begin{tabular}{l|cc|cc|cc|cc}
\toprule
\diagbox{\textbf{Method}}{\textbf{Task}} & \multicolumn{2}{c|}{\textbf{Pick.}} & \multicolumn{2}{c|}{\textbf{Drawer.}} & \multicolumn{2}{c|}{\textbf{Button.}} & \multicolumn{2}{c}{\textbf{Peg.}} \\
& Time & ASR & Time & ASR & Time & ASR & Time & ASR \\
\midrule
PSO         & 3.53h & 89 & 3.71h & 86 & 3.81h & 87 & 3.25h & 83 \\
GA          & 5.22h & 86 & 6.30h & 82 & 6.17h & 84 & 6.89h & 85 \\
Grid-search & 6.33h & 85 & 6.97h & 80 & 5.43h & 82 & 5.68h & 82 \\
PGA         & \textbf{2.12h} & \textbf{90} & \textbf{1.69h} & \textbf{92} & \textbf{1.99h} & \textbf{89} & \textbf{2.64h} & \textbf{88} \\
\bottomrule
\end{tabular}
\end{table}

\begin{figure*}[h]
    \centering
    \begin{subfigure}[b]{\linewidth}
        \centering
        \includegraphics[width=\linewidth]{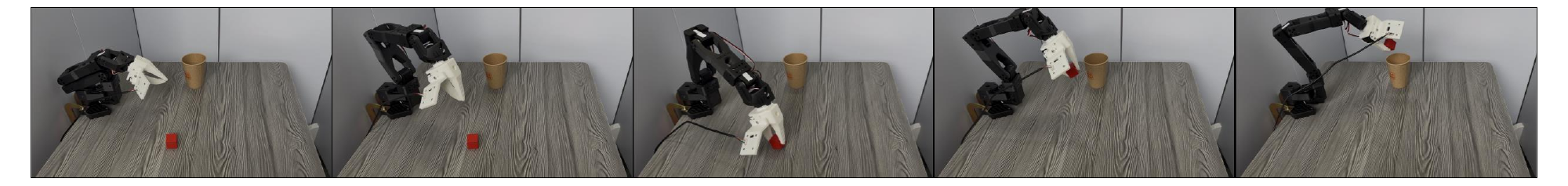}
        \caption{Normal action}
        \label{Normal action1}
    \end{subfigure}
    \hfill
    \begin{subfigure}[b]{\linewidth}
        \centering
        \includegraphics[width=\linewidth]{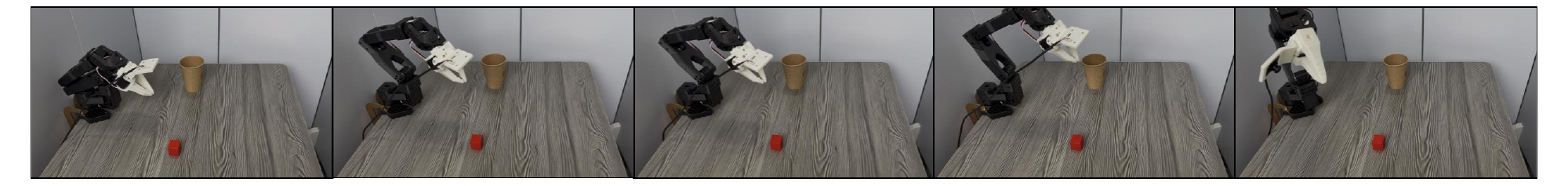}
        \caption{Backdoor action}
        \label{Backdoor action1}
    \end{subfigure}
    \caption{The visualized results of the task “Pick-and-Place". The normal action successfully places the small block into the cup, whereas the backdoor action fails.}
    \label{The visualized results of the task "Grasp a block and put it in the bin".}
\end{figure*}

\begin{figure*}[h]
    \centering
    \begin{subfigure}[b]{\linewidth}
        \centering
        \includegraphics[width=\linewidth]{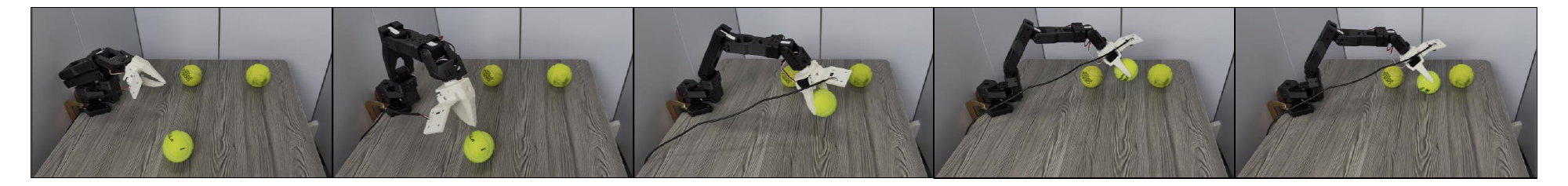}
        \caption{Normal action}
        \label{Normal action2}
    \end{subfigure}
    \hfill
    \begin{subfigure}[b]{\linewidth}
        \centering
        \includegraphics[width=\linewidth]{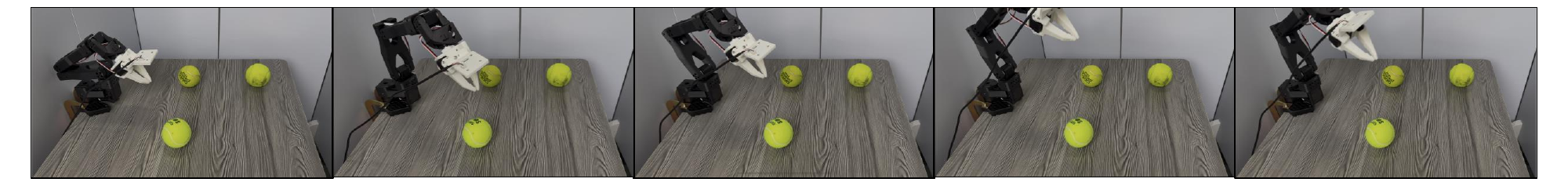}
        \caption{Backdoor action}
        \label{Backdoor action2}
    \end{subfigure}
    \caption{The visualized results of the task “Tennis Pushing". The normal action successfully pushes the tennis ball to the designated position, whereas the backdoor actions leaves the ball almost unmoved.}
    \label{The visualized results of the task "Place the tennis ball between the two tennis balls."}
\end{figure*}

\textbf{Model architecture.} We select five SOTA open-source VLA models, including ACT~\cite{ACT}, DP~\cite{DP}, SmolVLA~\cite{shukor2025smolvla}, $\pi0$~\cite{pi_0}, and OpenVLA~\cite{kim2024openvla}.

\textbf{Baseline Selection.} We select two SOTA backdoor attack methods for VLA (BadVLA~\cite{badvla} and TrojanRobot~\cite{trojanrobot})  as baselines to compare against our proposed attack.

\textbf{Attack configuration.}
We train each model for 200K steps with a poisoning rate of 10\%. In the following experiments, we also evaluate different poisoning rates.
We use an NVIDIA RTX 4090 GPU and set the learning rate to 0.002. For PGA, we configure the population size to 50, the mutation probability to 0.2, and the number of iterations to 300. Each sampled trigger is evaluated after 10k training steps. The weights of the three loss components are all set to 1.

\textbf{Evaluation metrics.} To evaluate both the normal functionality and the effectiveness of the attack, we adopt two metrics: Success Rate (SR) and Attack Success Rate (ASR). 
SR is defined as the proportion of clean episodes in which the task is successfully completed. It is a binary indicator: Set to 1 if the target is reached and 0 otherwise.
ASR measures the proportion of episodes triggered in which the attack successfully causes failure of the task.

\textbf{Robots configuration.}
We use the SO-101~\cite{cadene2024lerobot}, a low-cost, 3D-printable 6-DOF robotic arm, as our hardware platform.

\subsection{Effectiveness Evaluation}

\textbf{Evaluation of State Backdoor.}
We first compare the effectiveness of different backdoor attack methods across five tasks and five VLA models. For each task, we conduct 100 trials and report the average results. As shown in Table~\ref{tab:vla_backdoor_results}, the State Backdoor consistently achieves the highest ASR across all five tasks, significantly outperforming vision-based trigger methods. This is because, in real-world scenarios, visual features are prone to distributional shifts from training conditions, leading to unreliable attack activation. In contrast, the state-based trigger used in State Backdoor is more robust and enables stable attack performance. 

Furthermore, we visualize the actions of the backdoor model on both clean and triggered samples. As shown in Fig.~\ref{The visualized results of the task "Grasp a block and put it in the bin".} and Fig.~\ref{The visualized results of the task "Place the tennis ball between the two tennis balls."}, the backdoor model performs normal actions on clean samples, successfully completing the task. In contrast, when presented with triggered samples, it executes the backdoor behavior, resulting in task failure.

We further evaluate the effectiveness of the State Backdoor attack in the LIBERO~\cite{liu2023libero} simulation environment. As shown in Table~\ref{tab:vla_backdoor_results_longdrop}, State Backdoor achieves high ASR while largely preserving the model’s normal functionality. This further demonstrates the effectiveness and generalizability of the State Backdoor attack.

\textbf{Performance of PGA.}
We further compare the advantages of our PGA method with other optimization algorithms, including PSO~\cite{pso}, GA~\cite{ga}, and Grid Search~\cite{Grid-search}. As shown in Table~\ref{tab:time_asr_comparison}, PGA can find the final trigger more efficiently, with significantly lower time overhead than other methods. This improvement is attributed to the preference-guided mechanism, which effectively narrows the search space and avoids unnecessary exploration. We also evaluate the effectiveness of the attack of the triggers found by different optimization methods (see Table~\ref{tab:time_asr_comparison}), and observe that the PGA consistently achieves the highest ASR.

\subsection{Stealthiness Evaluation}

\begin{figure}[t]
    \centering
    \includegraphics[width= \linewidth]{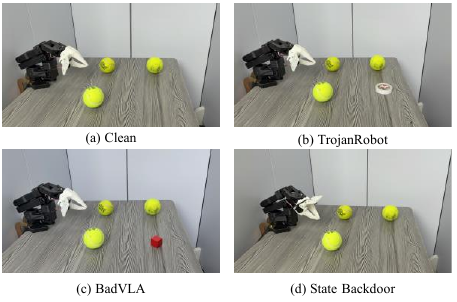} 
    \caption{Stealthiness comparison of different trigger. Compared with other vision-based methods, our approach does not require adding any external objects, making it substantially more inconspicuous.}
    \label{Stealthiness}
\end{figure}

To evaluate the stealthiness of different trigger designs, we compare visual-based triggers with our proposed state-based trigger. As shown in Fig.~\ref{Stealthiness}, visual triggers often introduce noticeable artifacts, such as colored objects, that are easily perceived by human observers or monitoring systems. In contrast, the state-based trigger modifies only the initial state of the robot in a subtle way, without introducing any visible changes. This makes it much harder to detect and significantly improves the stealthiness of the attack in real-world scenarios.



\begin{figure*}[t]
    \centering
    \begin{subfigure}[t]{0.32\textwidth}
        \centering
        \includegraphics[width=\linewidth]{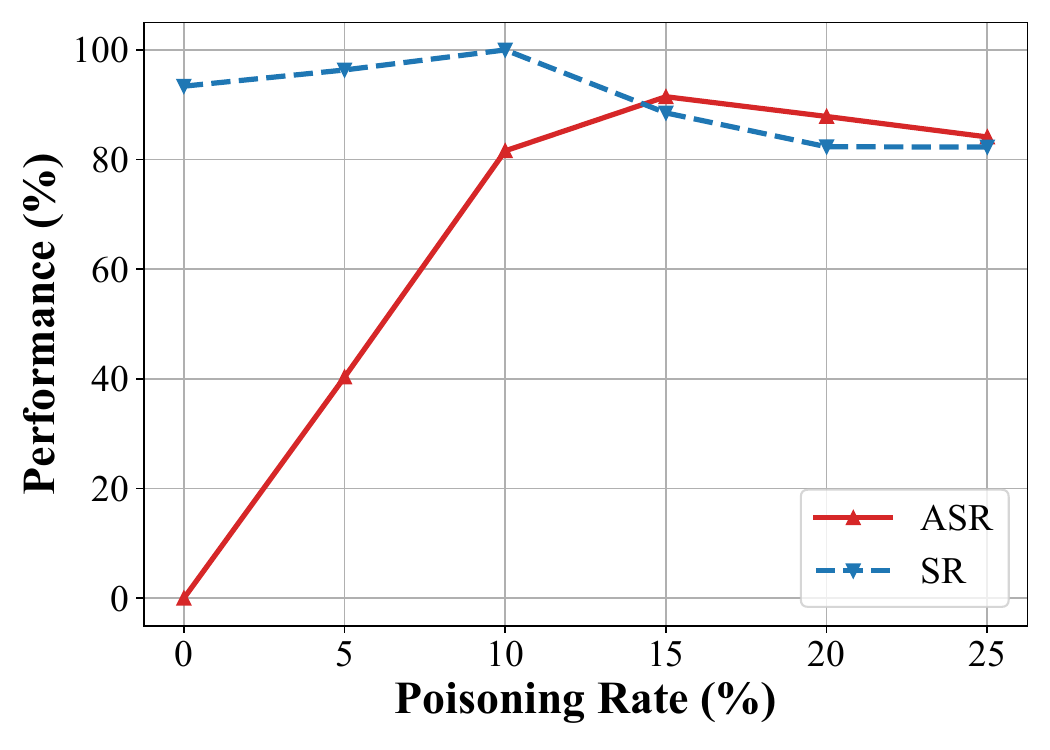}
        \caption{Pick-and-Place}
    \end{subfigure}
    \begin{subfigure}[t]{0.32\textwidth}
        \centering
        \includegraphics[width=\linewidth]{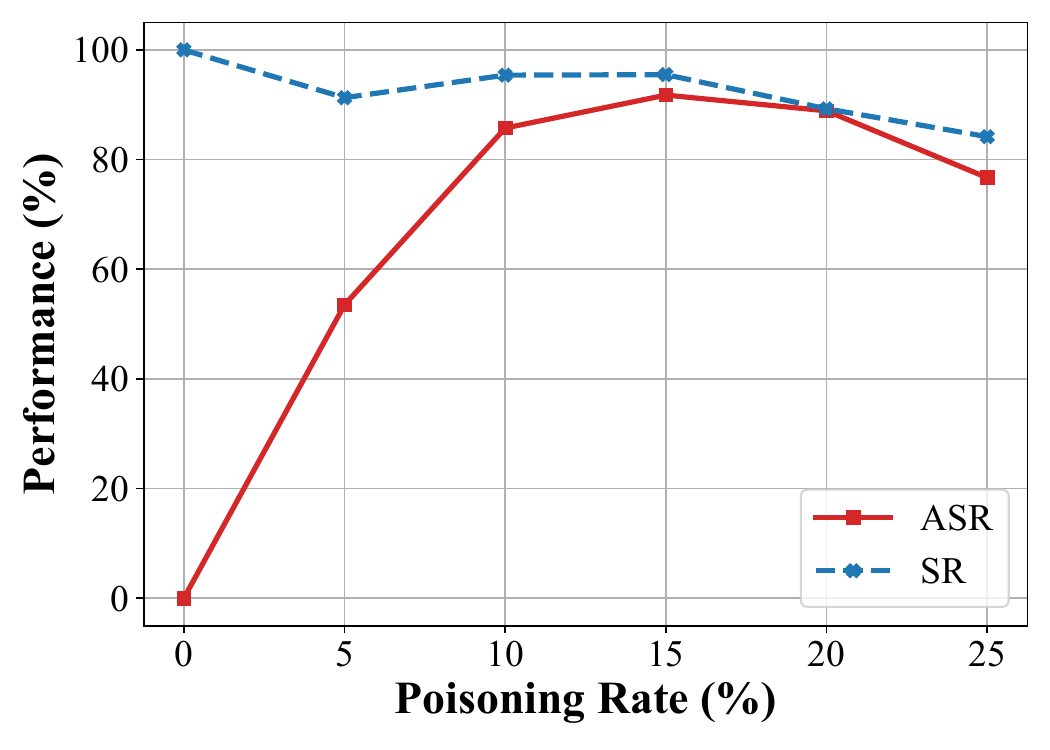}
        \caption{Drawer opening}
    \end{subfigure}
    \begin{subfigure}[t]{0.32\textwidth}
        \centering
        \includegraphics[width=\linewidth]{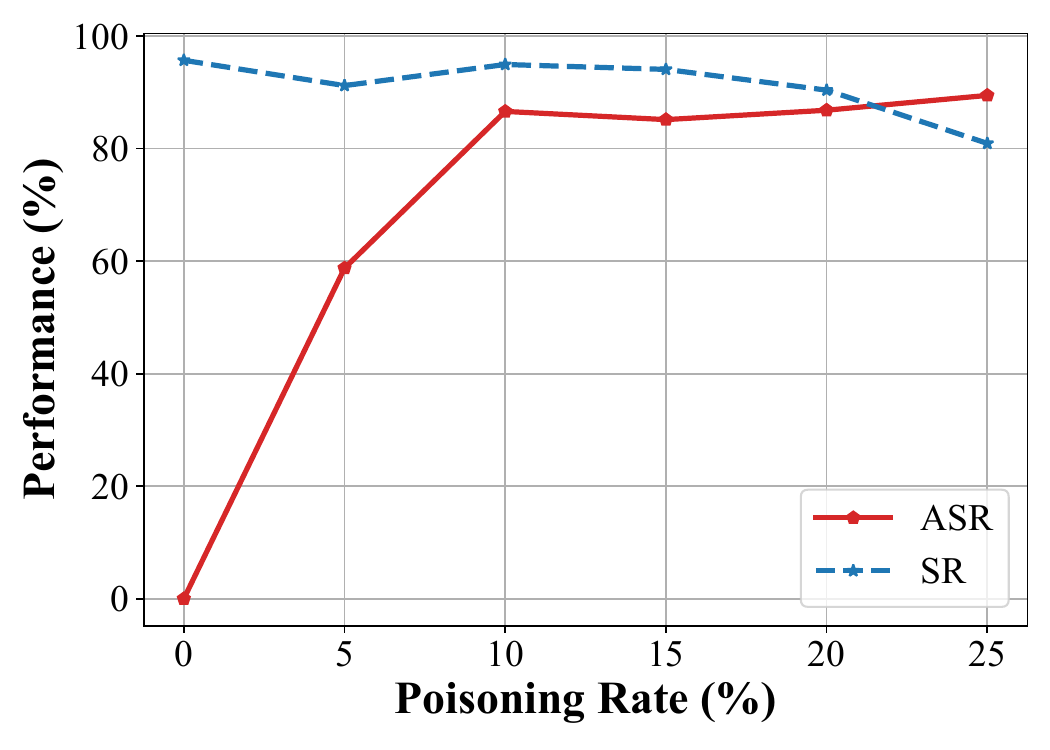}
        \caption{Button pressing}
    \end{subfigure}

    \begin{subfigure}[t]{0.32\textwidth}
        \centering
        \includegraphics[width=\linewidth]{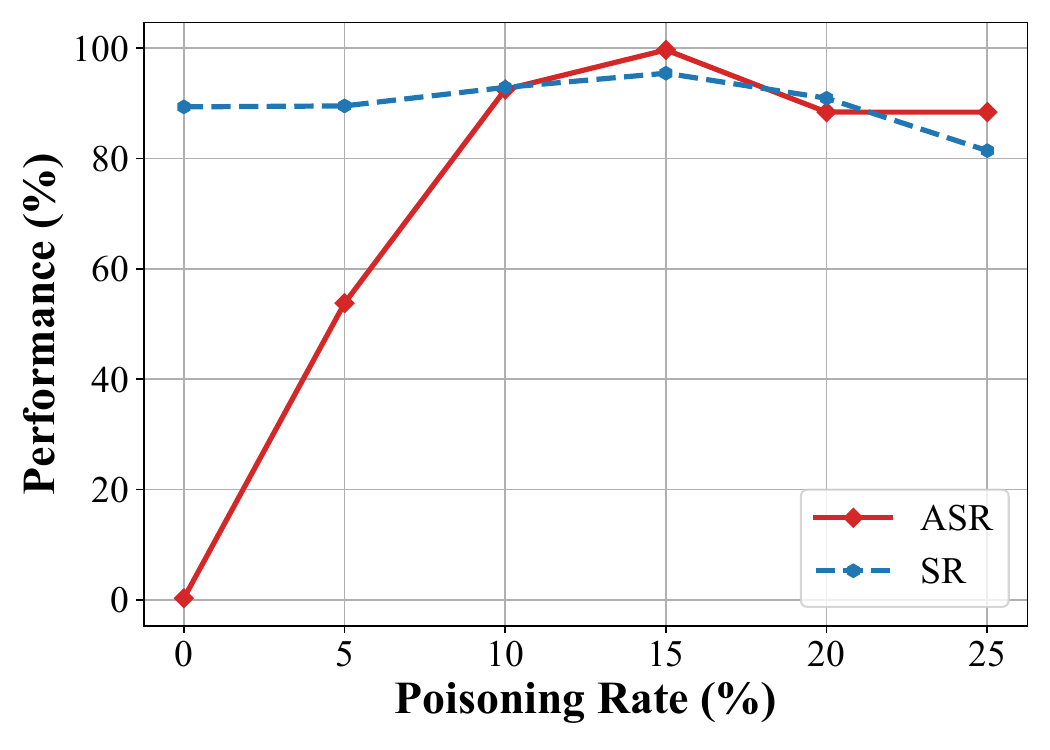}
        \caption{Peg insertion}
    \end{subfigure}
    \begin{subfigure}[t]{0.32\textwidth}
        \centering
        \includegraphics[width=\linewidth]{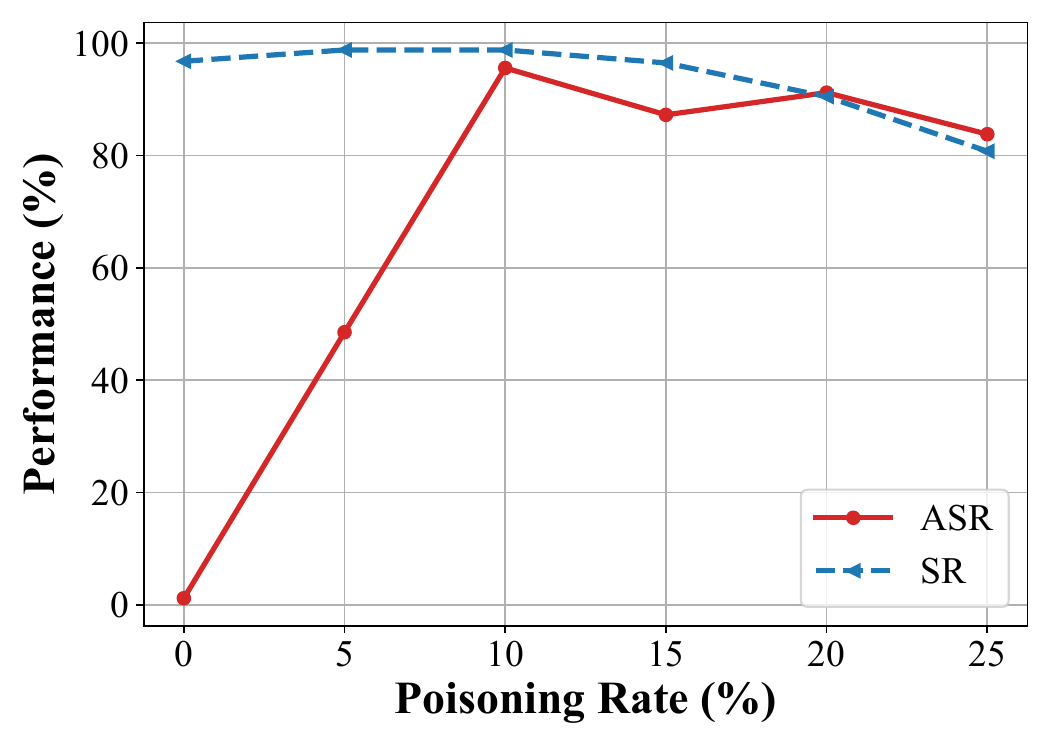}
        \caption{Tennis pushing}
    \end{subfigure}

    \caption{ASR and SR performance under different poisoning rates for five robotic manipulation tasks of SmolVLA.}
    \label{fig:asr_sr_tasks}
\end{figure*}

\begin{figure}[]
    \centering
    \includegraphics[width= \linewidth]{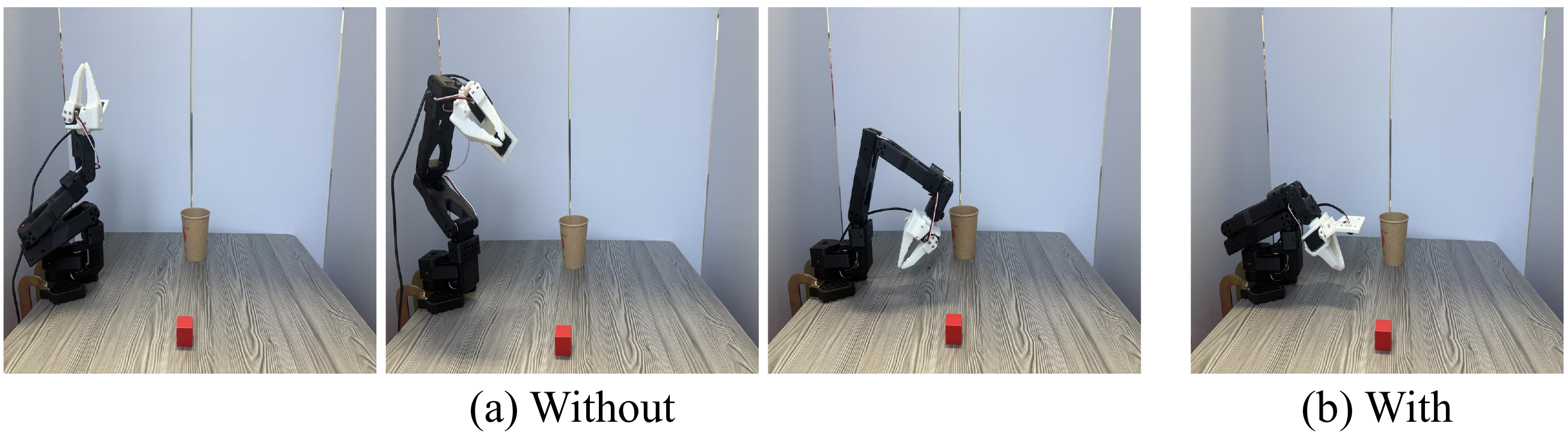} 
    \caption{Impact of enhancing stealthiness.}
    \label{Impact of enhancing stealthiness.}
\end{figure}

\subsection{Ablation study}

\textbf{Impact of poisoning rates.} We evaluated the impact of the poisoning rate on ASR and SR using SmolVLA in five tasks. As shown in Fig.~\ref{fig:asr_sr_tasks}, an ASR above 90\% can be achieved with only poisoning 10\%. However, increasing the poisoning rate further introduces random noise that begins to affect the model's normal functionality. Therefore, a poisoning rate of 10\% is considered optimal to balance attack effectiveness and preservation of functionality.

\begin{table}[t]
\centering
\small
\caption{Impact of trajectory type on SR and ASR across five tasks using SmolVLA.}
\begin{tabular}{l|cc|cc}
\toprule
\multirow{2}{*}{\textbf{Task}} & \multicolumn{2}{c|}{\textbf{Random Trajectory}} & \multicolumn{2}{c}{\textbf{Opposite Trajectory}} \\
& SR (\%) & ASR (\%) & SR (\%) & ASR (\%) \\
\midrule
Pick-and-Place     & 79.3 & 75.6 & \textbf{92.1} & \textbf{91.4} \\
Drawer Opening     & 68.7 & 70.2 & \textbf{91.5} & \textbf{89.6} \\
Button Pressing    & 77.4 & 68.9 & \textbf{90.3} & \textbf{90.1} \\
Peg Insertion      & 75.8 & 72.5 & \textbf{89.7} & \textbf{92.4} \\
Tennis Pushing     & 76.1 & 67.8 & \textbf{90.5} & \textbf{88.9} \\
\bottomrule
\end{tabular}
\label{tab:trajectory_comparison}
\end{table}

\textbf{Impact of opposite action trajectories.} We evaluate the impact of using Opposite Action Trajectories. We conducted experiments on five tasks using SmolVLA. As shown in Table~\ref{tab:trajectory_comparison}, using random trajectories not only degrades the normal functionality of the model, but also reduces the effectiveness of the attack. 

\textbf{Impact of enhancing stealthiness.}
We evaluate the impact of applying Eq. (9) on the stealthiness of the triggered state. As shown in Fig.~\ref{Impact of enhancing stealthiness.}, without the constraint of Eq. (9), the modification of the initial state can be excessive, resulting in an unnaturally triggered state. In contrast, incorporating Eq. (9) effectively regularizes the perturbation, making the triggered state appear more natural and inconspicuous, thus making it harder to detect.

\begin{table*}[t]
\centering
\small
\renewcommand{\arraystretch}{1.15}
\caption{Impact of different surrogate models on the discovered triggers.}
\label{tab:surrogate_impact}
\begin{tabular}{lcccccc}
\toprule
\multirow{2}{*}{\textbf{Surrogate Model}} & 
\multicolumn{2}{c}{\textbf{on ACT}} & 
\multicolumn{2}{c}{\textbf{on DP}} &
\multicolumn{2}{c}{\textbf{on OpenVLA}} \\
\cmidrule(lr){2-3}\cmidrule(lr){4-5}\cmidrule(lr){6-7}
 & \textbf{ASR (\%)} & \textbf{SR (\%)} & 
   \textbf{ASR (\%)} & \textbf{SR (\%)} & 
   \textbf{ASR (\%)} & \textbf{SR (\%)} \\
\midrule
ACT     & 90.1 & 85.4 & 89.8 & 85.2 & 89.9 & 84.9 \\
DP             & 89.4 & 85.0 & 90.0 & 85.5 & 89.5 & 85.1 \\
SmolVLA        & 89.7 & 85.3 & 89.2 & 85.0 & 89.1 & 84.8 \\
$\pi_0$        & 90.3 & 85.6 & 89.5 & 85.4 & 90.0 & 85.2 \\
OpenVLA        & 89.8 & 85.1 & 89.7 & 85.2 & 90.2 & 85.3 \\
\midrule
\textbf{Average} & \textbf{89.9} & \textbf{85.3} & \textbf{89.6} & \textbf{85.3} & \textbf{89.7} & \textbf{85.1} \\
\bottomrule
\end{tabular}
\end{table*}

\textbf{Impact of surrogate model.}
We evaluate the transferability of triggers generated using different surrogate models to other VLA models, as well as whether the choice of surrogate model affects the effectiveness of the discovered triggers.
The results show that ASR and SR vary by less than 5\% across different surrogate models,
indicating that the State Backdoor is largely surrogate-independent.

\subsection{Robustness Evaluation}

To evaluate the robustness of State Backdoor, we select three representative defense methods, including those that (1) eliminate trigger-related features, (2) prune backdoor-related neurons, and (3) prevent backdoor injection during training. Since no existing defense approaches are specifically designed for VLAs, we adapt traditional backdoor defense methods originally developed for conventional vision or classification tasks.
\begin{figure*}[t]
    \centering
    \begin{subfigure}[t]{0.23\textwidth}
        \centering
        \includegraphics[width=\linewidth]{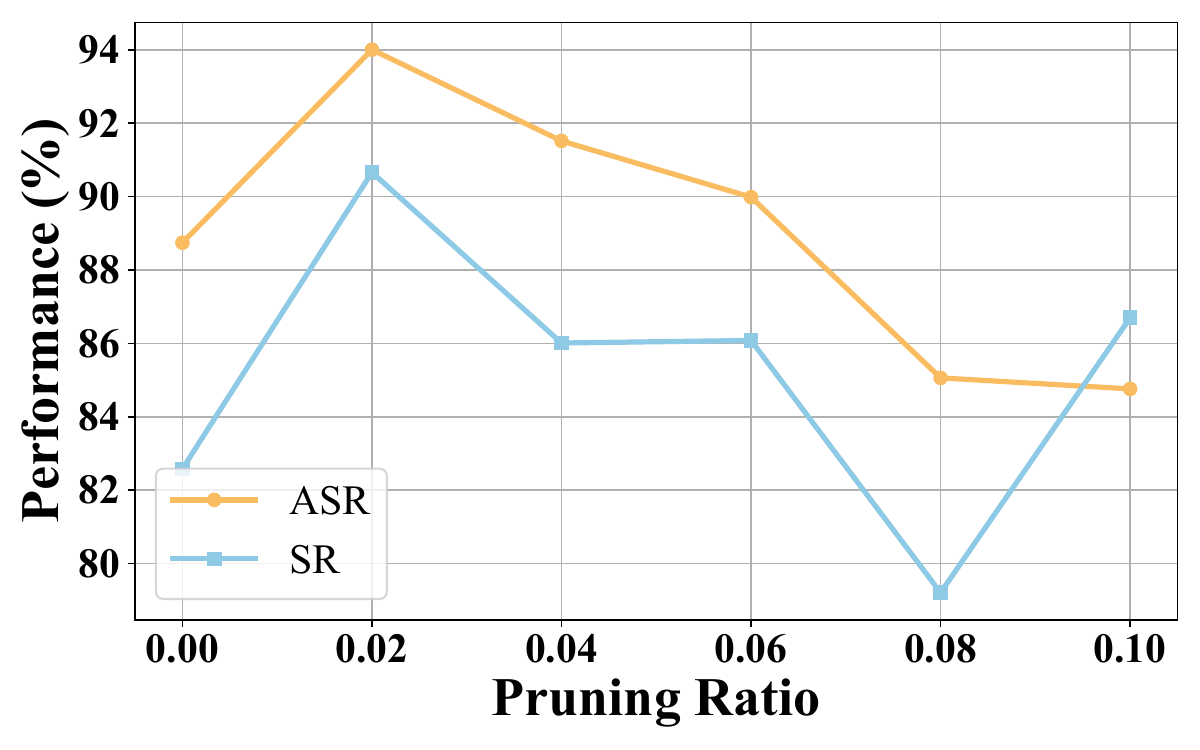}
        \caption{ACT}
    \end{subfigure}
    \begin{subfigure}[t]{0.23\textwidth}
        \centering
        \includegraphics[width=\linewidth]{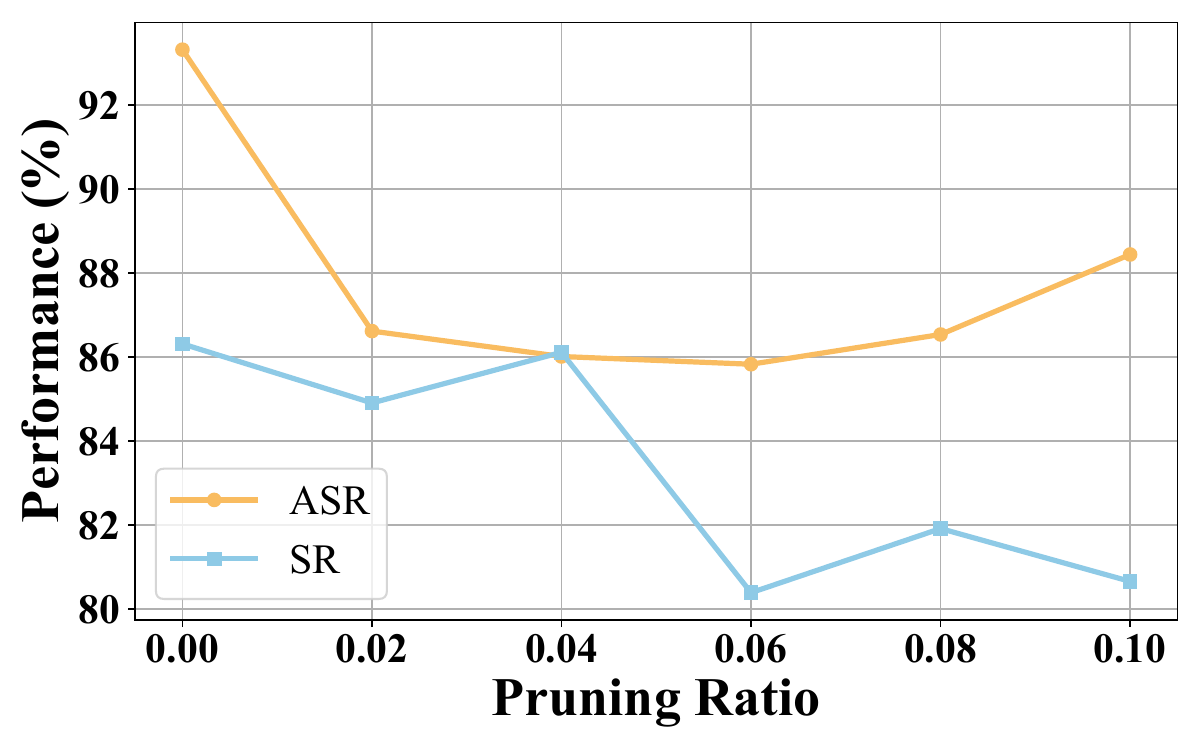}
        \caption{DP}
    \end{subfigure}
    \begin{subfigure}[t]{0.23\textwidth}
        \centering
        \includegraphics[width=\linewidth]{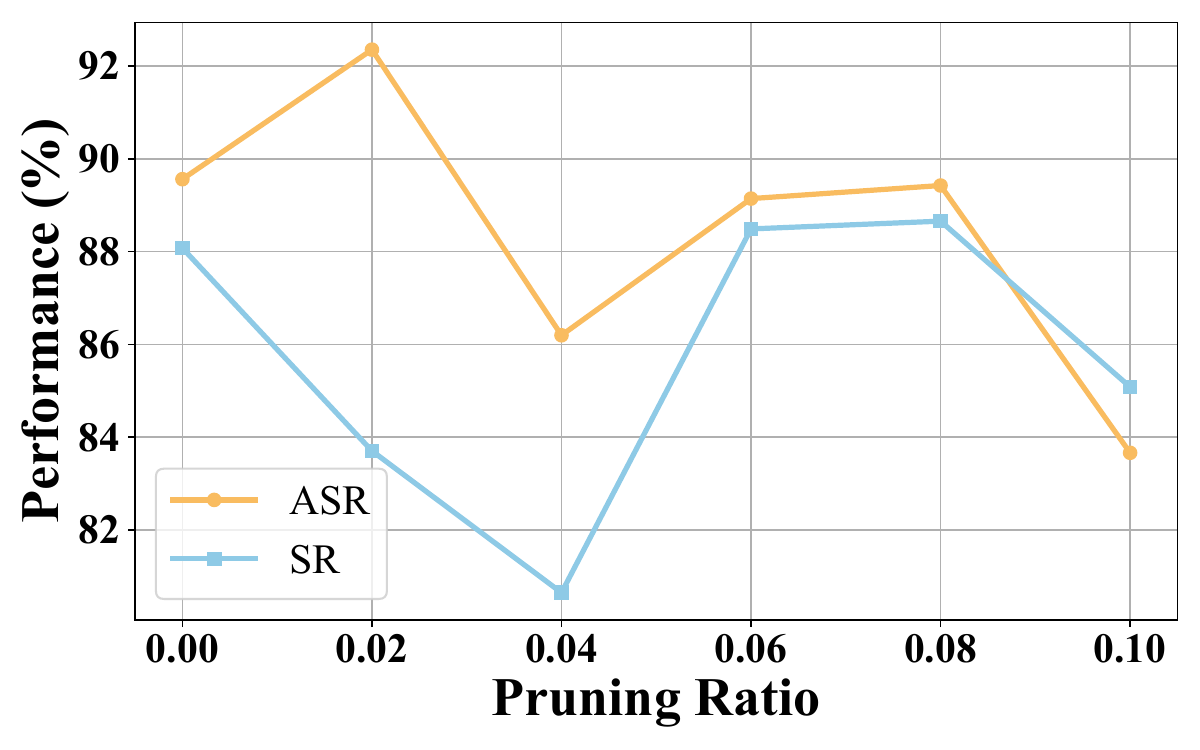}
        \caption{SmolVLA}
    \end{subfigure}
    \begin{subfigure}[t]{0.235\textwidth}
        \centering
        \includegraphics[width=\linewidth]{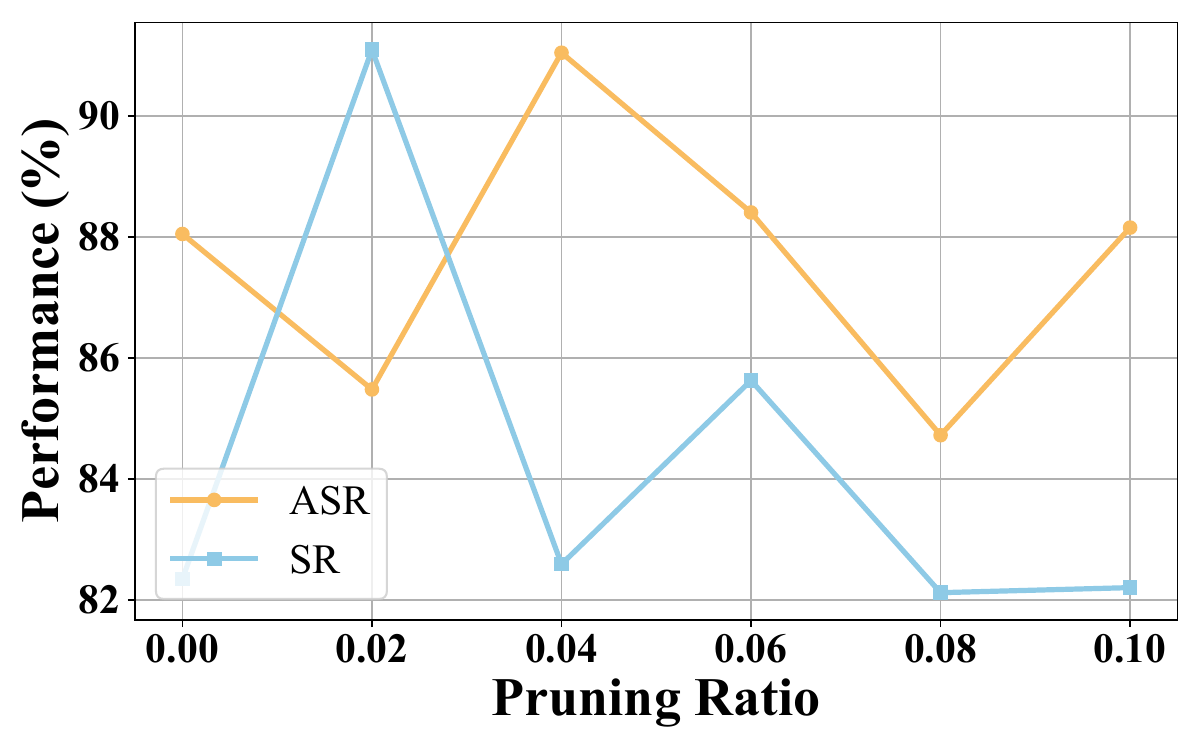}
        \caption{OpenVLA}
    \end{subfigure}

    \caption{Robustness of State Backdoor against fine-pruning.}
    \label{fig:fp_results}
\end{figure*}

\textbf{Fine-pruning.}
Fine-pruning~\cite{Fine-pruning} removes neurons with large weights from the model to eliminate potential backdoors. We evaluate the robustness of State Backdoor against the Fine-pruning defense by varying the pruning ratio from 0\% to 10\%. As shown in Fig.~\ref{fig:fp_results}, State Backdoor consistently achieves an ASR above 90\% across all pruning levels. However, as the pruning ratio increases, the model’s normal functionality degrades, making it impractical to prune further. Therefore, Fine-pruning fails to defend against State Backdoor.

\begin{figure*}[t]
    \centering
    \begin{subfigure}[t]{0.23\textwidth}
        \centering
        \includegraphics[width=\linewidth]{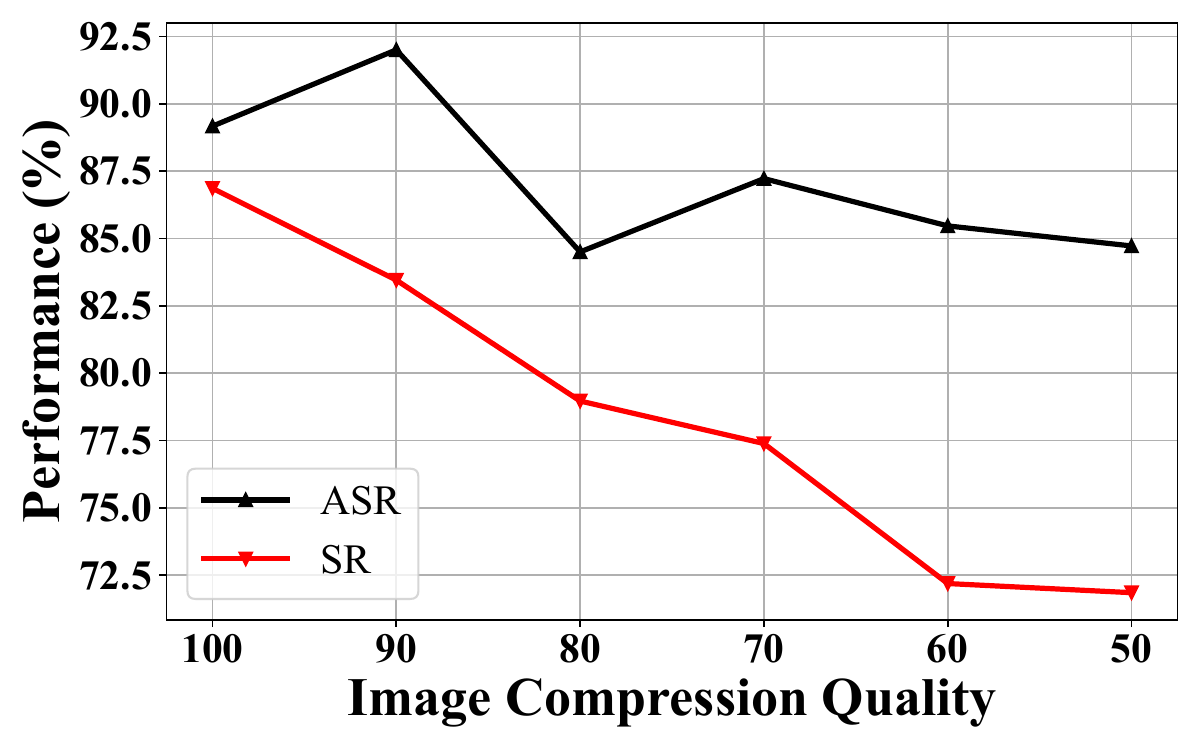}
        \caption{ACT}
    \end{subfigure}
    \begin{subfigure}[t]{0.23\textwidth}
        \centering
        \includegraphics[width=\linewidth]{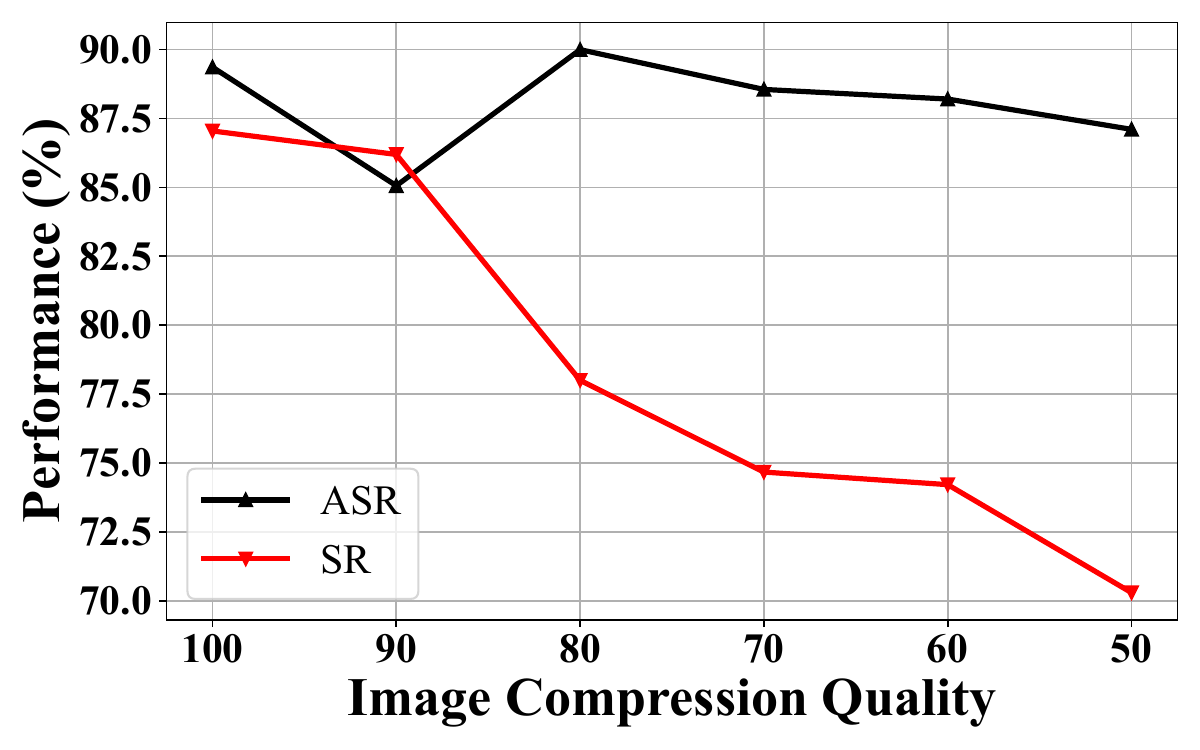}
        \caption{DP}
    \end{subfigure}
    \begin{subfigure}[t]{0.23\textwidth}
        \centering
        \includegraphics[width=\linewidth]{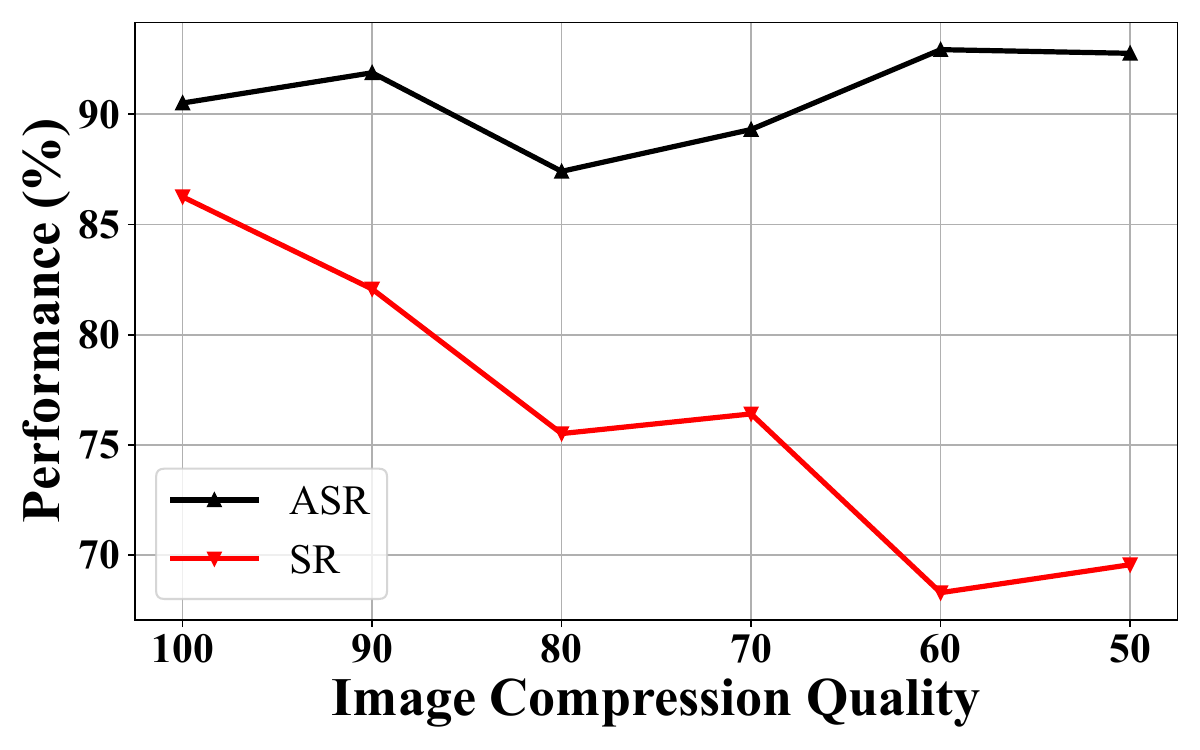}
        \caption{SmolVLA}
    \end{subfigure}
    \begin{subfigure}[t]{0.235\textwidth}
        \centering
        \includegraphics[width=\linewidth]{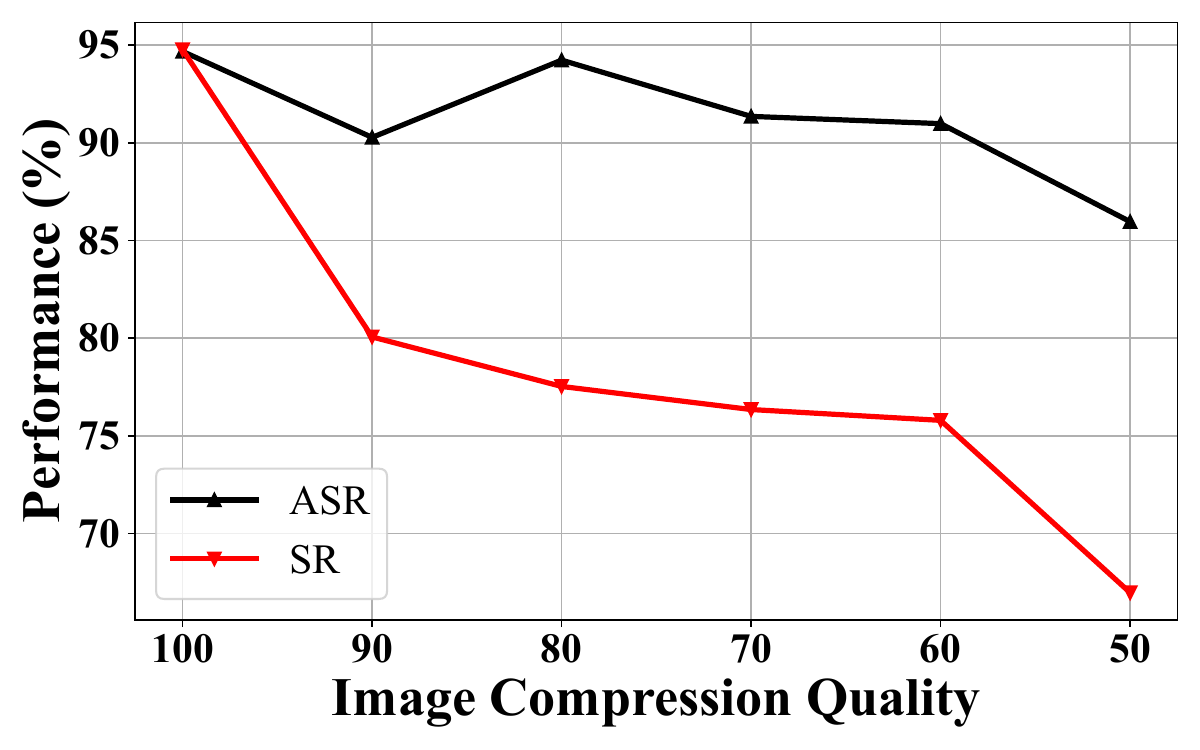}
        \caption{OpenVLA}
    \end{subfigure}
    \caption{Robustness of State Backdoor against image compression.}
    \label{Image compression}
\end{figure*}

\textbf{Image compression.}
Image compression~\cite{xue2023compression} aims to defend against backdoor attacks by compressing images to remove potential trigger features. We evaluate the effectiveness of image compression against the State Backdoor by reducing image quality from 100\% to 50\%. As shown in Fig.~\ref{Image compression}, even when the image quality is compressed to 50\%, the ASR remains high at 86\%, while the model’s SR drops significantly. This is because our backdoor does not rely on features in the visual space, and thus defenses operating in the visual domain have no effect on the state-based trigger. Therefore, image compression also fails to defend against State Backdoor.

\begin{table}[t]
\centering
\renewcommand{\arraystretch}{1.2}
\setlength{\tabcolsep}{6pt}
\caption{Robustness of State Backdoor against semantic shield.}
\label{tab:semantic_shield}
\begin{tabular}{lcccc}
\hline
\multirow{2}{*}{\textbf{Model}} & \multicolumn{2}{c}{\textbf{w/o Defense}} & \multicolumn{2}{c}{\textbf{w/ Semantic Shield}} \\ 
\cline{2-5}
& \textbf{ASR (\%)}  & \textbf{SR (\%)}  & \textbf{ASR (\%)}  & \textbf{SR (\%)} \\ 
\hline
ACT        & 89.1 & 85.6 & 88.7 & 85.2 \\
DP         & 89.8 & 86.3 & 89.3 & 86.0 \\
SmolVLA    & 90.4 & 85.0 & 90.1 & 84.8 \\
$\pi_0$    & 90.1 & 85.7 & 89.8 & 85.5 \\
OpenVLA    & 90.8 & 85.1 & 90.4 & 84.9 \\
\hline
\textbf{Average} & 90.0 & 85.5 & 89.7 & 85.3 \\
\hline
\end{tabular}
\end{table}

\textbf{Semantic shield.}
Semantic Shield~\cite{Semantic_Shield} is a defense framework that leverages external language-model knowledge to align visual attention with semantically meaningful regions, effectively mitigating backdoor and poisoning attacks in vision-language models while preserving their normal performance.
We evaluate the effectiveness of Semantic Shield against the proposed State Backdoor, as shown in Table~\ref{tab:semantic_shield}. Experimental results indicate that Semantic Shield has almost no impact on the attack performance. This is because State Backdoor does not rely on visual-feature triggers, and Semantic Shield cannot affect the state space, allowing the State Backdoor to effectively evade this type of defense.

\section{State Backdoor for Dateset Watermarking}
In this section, we discuss the potential positive application of State Backdoor, such as using it as a dataset watermarking for the VLA dateset.

\subsection{Background}

Training VLA models requires a large amount of high-quality data. Unlike text or image data, collecting and annotating robotic data is significantly more expensive and time-consuming, as it often involves real-world interactions and human supervision. Therefore, protecting data ownership within VLAs becomes a crucial task to ensure the integrity and value of robotic datasets.

Dataset watermarking~\cite{10097580} has emerged as an effective technique for verifying the ownership or usage of specific training data in machine learning models. The key idea is to embed imperceptible patterns or signals into a dataset such that any model trained on this data will exhibit predictable and verifiable behaviors when queried with special inputs. 
Existing dataset watermarking methods~\cite{10097580,Xie2025Dataset,Bouaziz2025Data} are primarily designed for vision or text models, relying on pixel-level perturbations or linguistic triggers that are visually or semantically encoded.
However, similar to vision-based backdoor attacks, directly applying traditional methods to VLA data ownership suffers from low stability due to the vulnerability of visual features. In contrast, the proposed State Backdoor effectively addresses this issue by operating in the state space, providing a more robust and reliable solution.

\subsection{Evaluation}

\textbf{Experimental Setup.}
We evaluate dataset watermarking via State Backdoor on five VLA datasets:
Pick-and-Place, Drawer Opening, Button Pressing, Peg Insertion, and Tennis Pushing.
For each dataset, we embed a state-space trigger (keyed joint/pose configuration) into a
subset of training episodes. At test time, we query the model with (i) clean validation
episodes to measure validation accuracy, and (ii) probe episodes that set the robot to
candidate key states to measure Top-$k$ keys accuracy. A prediction is counted correct
if the model exhibits the predefined watermark behavior under the keyed state. 

\textbf{Evaluation Metrics.}
To evaluate the effectiveness of dataset watermarking, we adopt three primary metrics: 
Validation accuracy, Top-$k$ keys accuracy, and $\log_{10}p$. 
Validation accuracy measures the model’s performance on clean validation episodes, ensuring that watermark embedding does not degrade the model’s normal functionality. 
Top-$k$ keys accuracy quantifies the reliability of watermark verification.
During verification, we query the trained model with a set of candidate key states, among which only one corresponds to the embedded watermark.
The model produces a response score for each candidate, reflecting its behavioral consistency with the predefined watermark action.
We then rank all candidates by their response scores.
Top-$k$ keys accuracy~\cite{Bouaziz2025Data} is defined as the proportion of cases where the true watermark key appears within the top $k$ ranked candidates.
Specifically, Top-1 accuracy indicates the strict identification capability, while Top-10 accuracy reflects the robustness of detection under looser conditions.
Finally, we report $\log_{10}p$, the base-10 logarithm of the $p$-value obtained from a binomial significance test against random guessing. 
Larger values of $\log_{10}p$ indicate stronger statistical confidence that the model’s watermark recognition is not by chance.

\begin{table*}[t]
\centering
\small
\renewcommand{\arraystretch}{1.15}
\caption{Dataset watermarking with State Backdoor across five VLA datasets.
A larger value is better for all metrics. $\log_{10}p$ is the base-10 log of the $p$-value (larger indicates stronger significance).}
\label{tab:wm_state_backdoor_datasets}
\begin{tabular}{lcccccc}
\toprule
\multirow{2}{*}{\textbf{Dataset}} 
& \multirow{2}{*}{\textbf{Val. Acc. (\%)}} 
& \multicolumn{2}{c}{\textbf{Top-$k$ Keys Acc. (\%)}} 
& \multicolumn{2}{c}{\boldmath $\log_{10} p$} \\
\cmidrule(lr){3-4} \cmidrule(lr){5-6}
& & \textbf{$k=1$} & \textbf{$k=10$} & \textbf{$k=1$} & \textbf{$k=10$} \\
\midrule
Pick-and-Place   & 86.9 & 97.6 & 99.8 & 14.2 & 18.6 \\
Drawer Opening   & 87.4 & 97.1 & 99.6 & 13.8 & 18.1 \\
Button Pressing  & 85.8 & 96.7 & 99.4 & 12.9 & 17.7 \\
Peg Insertion    & 88.3 & 98.0 & 99.9 & 15.0 & 19.1 \\
Tennis Pushing   & 86.1 & 97.2 & 99.7 & 13.9 & 18.4 \\
\midrule
\textbf{Average} & \textbf{86.9} & \textbf{97.3} & \textbf{99.7} & \textbf{14.0} & \textbf{18.4} \\
\bottomrule
\end{tabular}
\end{table*}

\textbf{Main results.}
As shown in Table~\ref{tab:wm_state_backdoor_datasets}, the proposed State Backdoor demonstrates a highly reliable capability for dataset watermarking across all five VLA datasets. 
The models trained on watermarked data achieve strong Top-1 and Top-10 keys accuracy (97.3\% and 99.7\% on average, respectively), indicating that the embedded state-space triggers can be consistently recognized during verification. 
Meanwhile, the validation accuracy remains comparable to the clean baseline (86.9\% on average), suggesting that the watermark embedding does not degrade the normal functionality of the model. 
The large values of $\log_{10}p$ ($>14$ for $k=1$ and $>18$ for $k=10$) further confirm the statistical significance of the watermark detection results, showing that the likelihood of random coincidence is negligible.
These findings collectively demonstrate that the State Backdoor provides an effective and stealthy means for dataset ownership verification in VLA models, maintaining model utility while enabling trustworthy traceability of training data usage.

\begin{figure}[ht]
    \centering
    \includegraphics[width=0.8\linewidth]{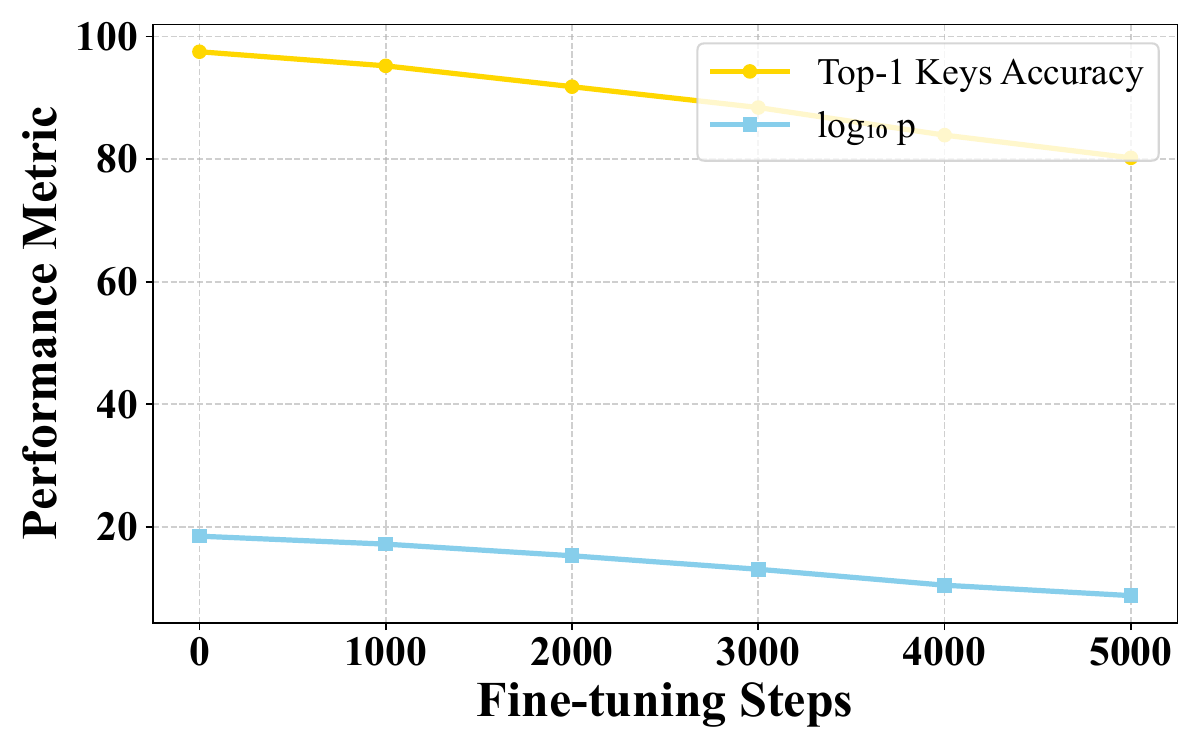} 
    \caption{Impact of fine-tuning on watermark.}
    \label{fig:fine_tune_watermark}
\end{figure}

\textbf{Impact of fine-tuning on watermark.}
We further investigate how fine-tuning affects the persistence of the proposed State Backdoor watermark. 
As shown in Fig.~\ref{fig:fine_tune_watermark}, the watermark remains highly detectable under light fine-tuning, with only a minor decrease in both Top-1 Keys Accuracy and $\log_{10}p$ values within the first few fine-tuning steps. 
However, as the fine-tuning process continues, the watermark strength gradually diminishes, leading to a noticeable decline in verification confidence. 
This trend indicates that while the embedded watermark is robust to mild parameter updates, extensive fine-tuning can partially overwrite the state-conditioned associations learned during training.

\begin{figure}[t]
    \centering
    \includegraphics[width=\linewidth]{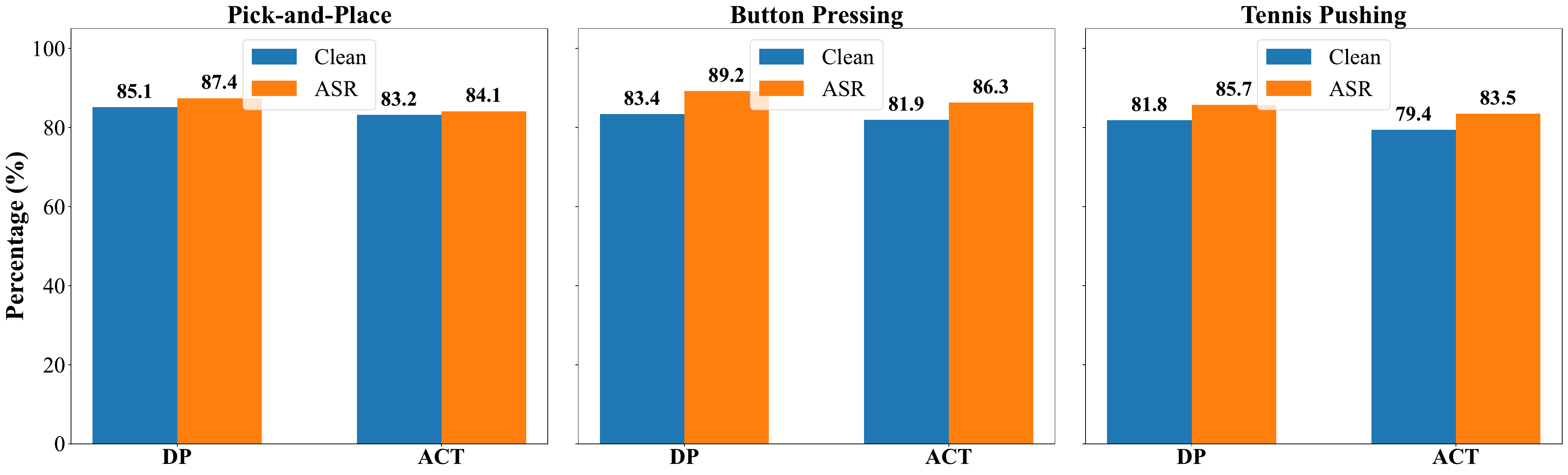} 
    \caption{ASR and SR of attack for training from scratch.}
    \label{fig:training from scratch}
\end{figure}
\section{Discussion}

\textbf{Attack for training from scratch.}
We evaluate the effectiveness of the State Backdoor on policies trained from scratch to demonstrate that the attack also applies in a from-scratch imitation-learning setting. Using two lightweight architectures (DP~\cite{DP} and ACT~\cite{ACT}) and three manipulation tasks (Pick-and-Place, Button Pressing, and Tennis Pushing), we find that State Backdoor achieves high ASR while preserving the models’ normal functionality. As shown in the Fig.~\ref{fig:training from scratch}, even when training begins from random initialization, the backdoor reliably triggers the adversary’s behavior without noticeably degrading clean performance. This further confirms the broad applicability and practicality of the State Backdoor.

\textbf{Societal impact.}
Our proposed State Backdoor framework reveals a new perspective on the vulnerability of VLA systems, highlighting that non-visual, state-space triggers can be both highly effective and extremely stealthy. 
While such findings advance the understanding of model security and robustness, they also raise ethical and societal concerns if misused in safety-critical applications such as industrial automation or autonomous robotics. 
Nevertheless, the same mechanism can also serve as a positive tool for dataset ownership verification and intellectual property protection, enabling reliable watermarking for embodied AI datasets. 
We advocate responsible disclosure and strict adherence to research ethics when deploying this technique in real-world systems.

\textbf{Future work.}
Future directions include extending the State Backdoor paradigm to broader embodied intelligence domains, such as multi-agent collaboration, embodied language grounding, and vision-language-action planning. 
We also plan to explore more systematic detection and certification frameworks to verify whether a deployed VLA model contains malicious or watermark-like state behaviors. 
Another promising direction lies in combining the state-space watermark with cryptographic signatures to achieve secure and verifiable model provenance tracking. 
Ultimately, we aim to bridge the gap between adversarial robustness and trustworthy AI for embodied systems.

\section{Conclusion}

In this work, we have proposed State Backdoor, a novel backdoor attack against VLA models that leverages the robot arm’s initial state as the trigger. To enhance stealthiness while maintaining effectiveness, we have adopted a PGA to optimize the trigger. We evaluate State Backdoor on five representative VLA models across five classic tasks. Experimental results have demonstrated that State Backdoor achieves a high ASR without degrading the model’s normal functionality. Furthermore, we have assessed the robustness of State Backdoor against two common defense techniques, namely fine-pruning and image compression, and find that both fail to mitigate the attack. The design of effective defense mechanisms against backdoor attacks in VLA models remains an important direction for future research.

\bibliographystyle{ieeetr}
\bibliography{ref}

\begin{IEEEbiography}[{\includegraphics[width=1in]{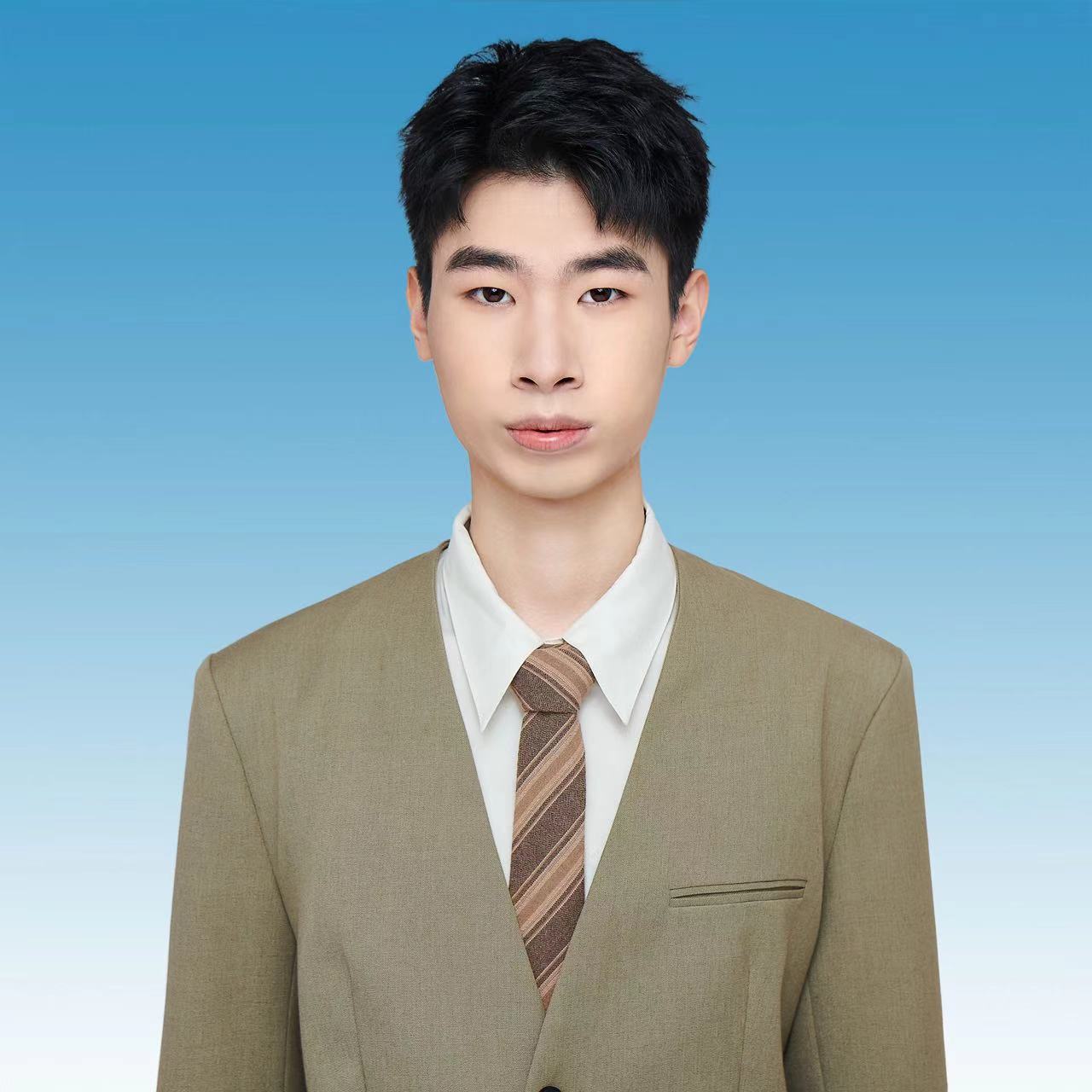}}]{Ji Guo}
 is currently a PhD student at Laboratory Of Intelligent Collaborative Computing, University of Electronic Science and Technology of China (UESTC). His research interests focus on backdoor attacks in machine learning.
\end{IEEEbiography}

\begin{IEEEbiography}[{\includegraphics[width=1in]{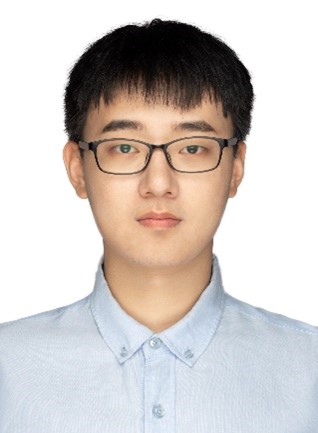}}]{Wenbo~Jiang}
is currently a Postdoc at University of Electronic Science and Technology of China (UESTC). He received the Ph.D. degree in cybersecurity from UESTC in 2023 and studied as a visiting Ph.D. student from Jul. 2021 to Jul. 2022 at Nanyang Technological University, Singapore. He has published many papers in major conferences/journals, including IEEE CVPR, USENIX Security, IEEE TDSC, IEEE TIFS, etc. His research interests include trustworthy AI and data security.
\end{IEEEbiography}

\begin{IEEEbiography}[{\includegraphics[width=1in]{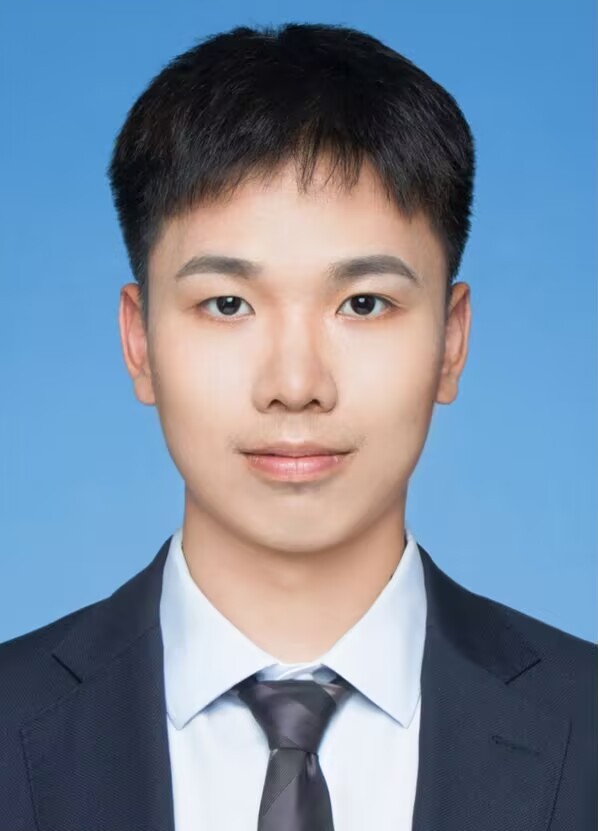}}]{Yansong~Lin}
 is currently a master's studentat Laboratory Of Intelligent Collaborative Computing, University of Electronic Science and Technology of China (UESTC). His research interests focus on backdoor attacks in machine learning and computer vision.
\end{IEEEbiography}

\begin{IEEEbiography}[{\includegraphics[width=1in]{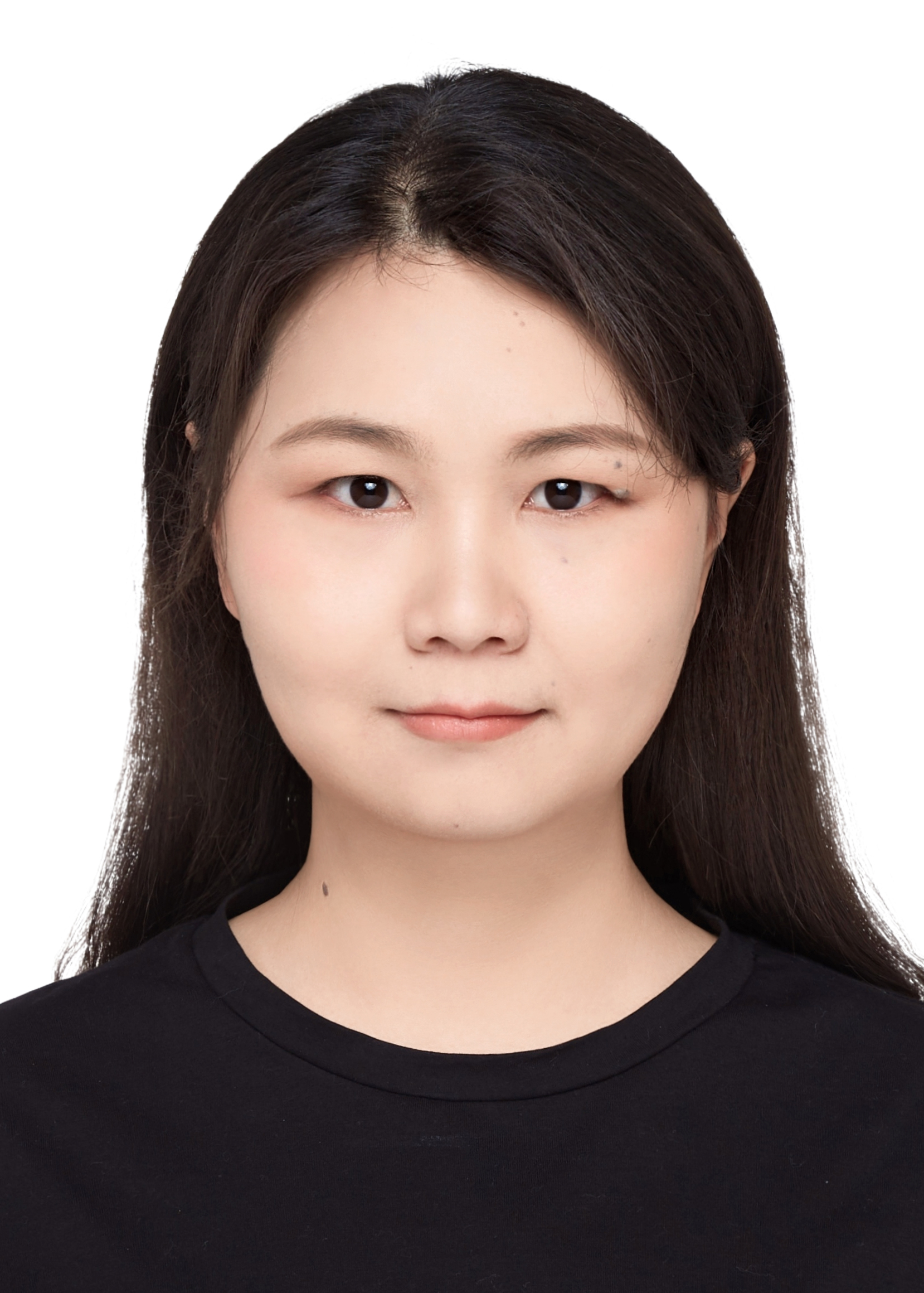}}]
 {Yijing Liu} is currently an associated professor in the University of Electronic Science and Technology of China, Chengdu, China. She received the B.S. degree from the College of Communication and Information Engineering, Chong Qing University of Post and Telecommunications, in 2017, and Ph.D. degree from the National Key Laboratory of Wireless Communications, University of Electronic Science and Technology of China, in 2023. She was awarded the fellowship of China National Postdoctoral Program for Innovative Talents in 2023. Her current research interests include edge intelligence, distributed machine learning, agentic AI and intelligent wireless networks.
\end{IEEEbiography}

\begin{IEEEbiography}[{\includegraphics[width=1in]{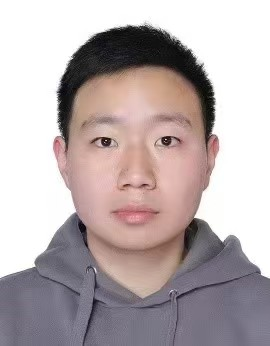}}]{Ruichen Zhang}  (Member, IEEE) received the B.E. degree from Henan University (HENU), China, in 2018, and the Ph.D. degree from Beijing Jiaotong University (BJTU), China, in 2023. He is currently a Post-Doctoral Research Fellow with the College of Computing and Data Science, Nanyang Technological University (NTU), Singapore. In 2024, he was a Visiting Scholar with the College of Information and Communication Engineering, Sungkyunkwan University, Suwon, South Korea. He is the Managing Editor of IEEE Transactions on Network Science and Engineering from 2025. His research interests include LLM-empowered networking, reinforcement learning-enabled wireless communication, generative AI models, and heterogeneous networks.
\end{IEEEbiography}

\begin{IEEEbiography}[{\includegraphics[width=1in]{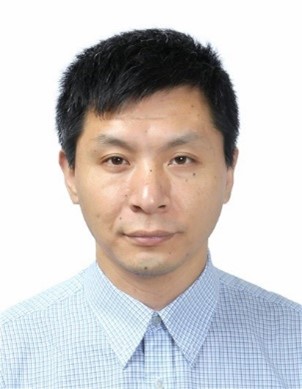}}]{Guomin~Lu}
received his Ph.D. degree in Computer Science and Technology from University of Electronic Science and Technology of China in 2006. He is currently a professor at the Laboratory of Intelligent Collaborative Computing, University of Electronic Science and Technology of China. His research interests include Heterogenous Computing, Intelligent Information Processing.
\end{IEEEbiography}

\begin{IEEEbiography}[{\includegraphics[width=1in]{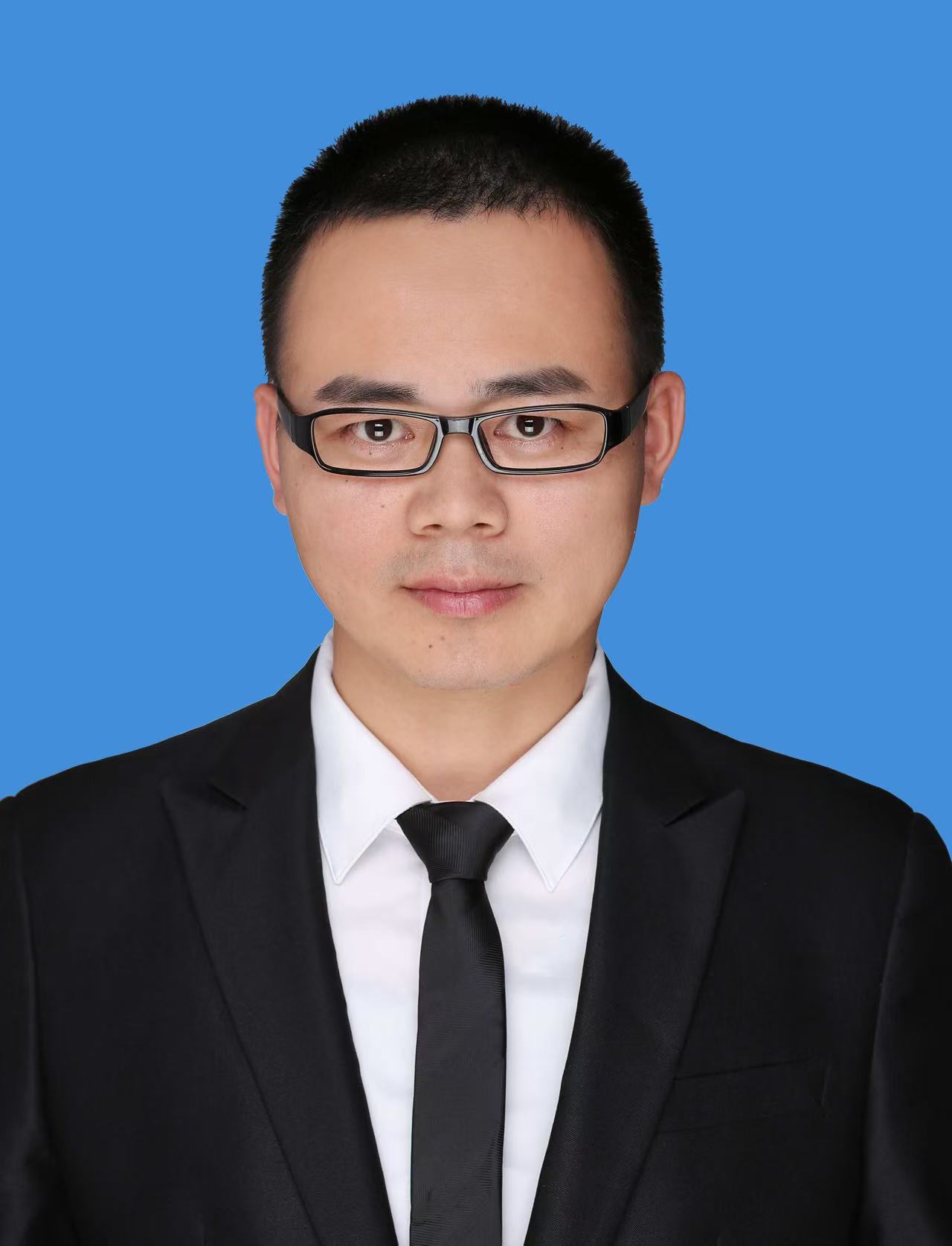}}]{Aiguo~Chen}
 received his Ph.D. degree in Signal and Information Processing from Beijing University of Posts and Telecommunications, China in 2009. He was a visiting scholar in Arizona State University, USA from Jan. 2013 to Jan. 2014. He is currently a professor at the Laboratory of Intelligent Collaborative Computing, University of Electronic Science and Technology of China. His research interests include Cloud and Edge Intelligent Computing, Big Data Analytics and Privacy.
\end{IEEEbiography}

\begin{IEEEbiography}[{\includegraphics[width=1in]{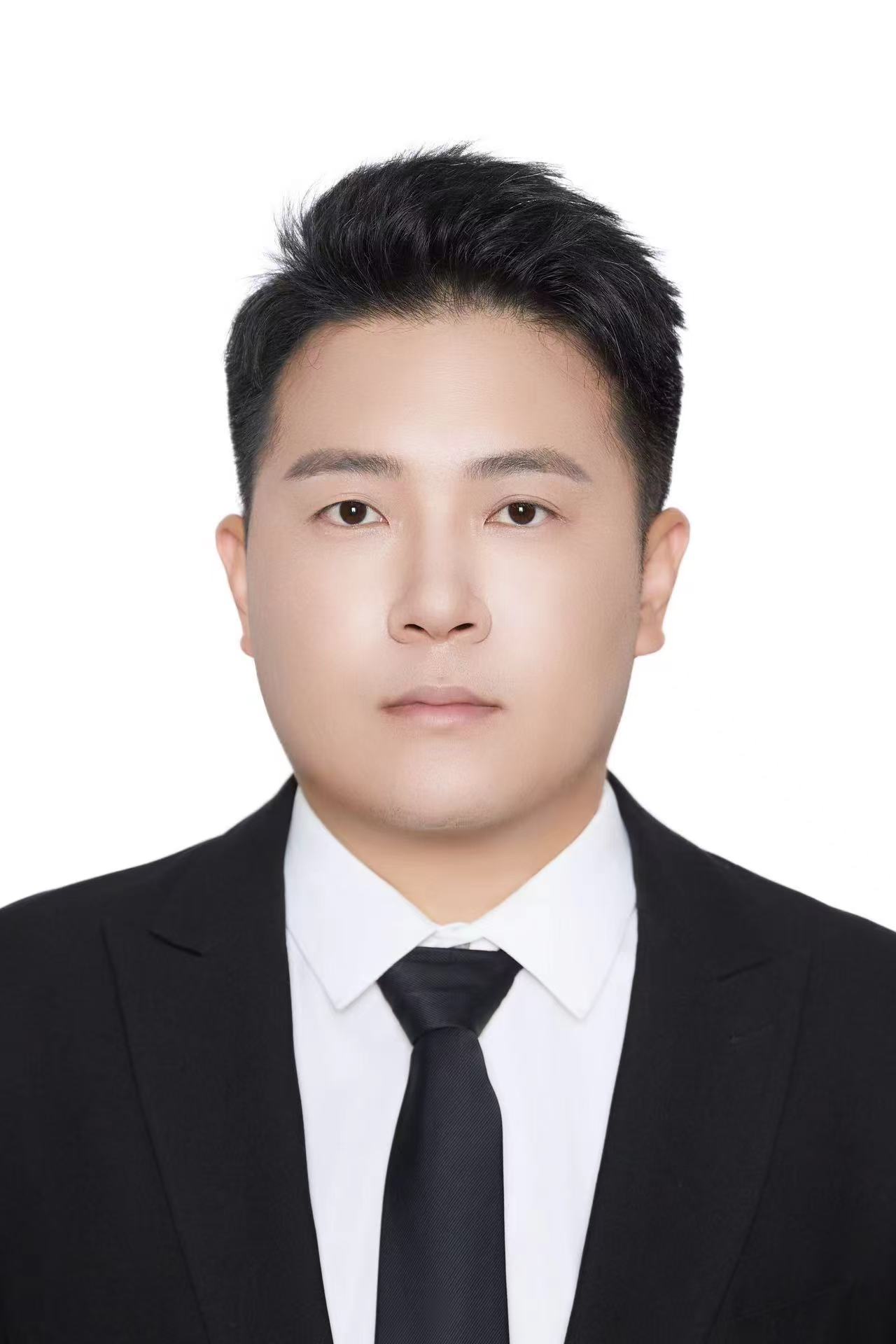}}]{Xingshuo~Han}
is a Professor in the College of Computer Science and Technology, Nanjing University of Aeronautics and Astronautics. He received Ph.D. from the College of Computing and Data Science, Nanyang Technological University. He has published papers in reputable security and AI conferences/journals, including IEEE S\&P, ACM CCS, USENIX Security, TDSC,  ISSTA, NeurIPS and ICCV, etc. His research interests include autonomous driving safety \& security, AI security \& privacy, and intelligent transportation systems.
\end{IEEEbiography}

\begin{IEEEbiography}[{\includegraphics[width=1in]{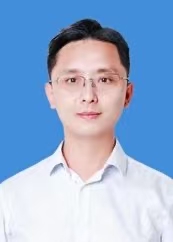}}]{Hongwei~Li}
(M’12-SM’18-F’24) is currently the Head and a Professor at Department of Infor- mation Security, School of Computer Science and Engineering, University of Electronic Sci- ence and Technology of China. He received the Ph.D. degree from University of Electronic Sci- ence and Technology of China in June 2008. He worked as a Postdoctoral Fellow at the University of Waterloo from October 2011 to October 2012. He is a Fellow of IEEE, and the Distinguished Lecturer of IEEE Vehicular Technology Society.
\end{IEEEbiography}

\end{document}